\documentclass[iop]{emulateapj}

\usepackage{graphicx}
\usepackage{amsmath}
\usepackage{ulem}

\newcommand{\ms}{\ifmmode{\rm m\thinspace s^{-1}}\else m\thinspace s$^{-1}$\fi}
\newcommand{\kms}{\ifmmode{\rm km\thinspace s^{-1}}\else km\thinspace s$^{-1}$\fi}
\newcommand{\kepler}{{\it Kepler\/}}

\newcommand{\vstar}{V482\;Per}

\shorttitle{\vstar}
\shortauthors{Torres et al.}

\begin{document}

\newcounter{aff}
\refstepcounter{aff}

\title{The quadruple-lined, doubly-eclipsing system V482 Persei}

\author{
Guillermo Torres\altaffilmark{1},
Claud H.\ Sandberg Lacy\altaffilmark{2},
Francis C.\ Fekel\altaffilmark{3},
Marek Wolf\altaffilmark{4}, \\ and
Matthew W.\ Muterspaugh\altaffilmark{3,5}
}

\altaffiltext{1}{Harvard-Smithsonian Center for Astrophysics, 60
  Garden Street, Cambridge, MA 02138, USA; gtorres@cfa.harvard.edu}

\altaffiltext{2}{Physics Department, University of Arkansas,
  Fayetteville, AR 72701, USA}

\altaffiltext{3}{Center of Excellence in Information Systems,
  Tennessee State University, Nashville, TN 37209, USA}

\altaffiltext{4}{Astronomical Institute, Faculty of Mathematics and
  Physics, Charles University in Prague, 180 00 Praha 8, Czech
  Republic}

\altaffiltext{5}{College of Life and Physical Sciences, Tennessee
  State University, Nashville, TN 37209, USA}

\begin{abstract}
We report spectroscopic and differential photometric observations of
the A-type system \vstar\ that reveal it to be a rare hierarchical
quadruple system containing two eclipsing binaries. One has the
previously known orbital period of 2.4 days and a circular orbit, and
the other a period of 6 days, a slightly eccentric orbit ($e = 0.11$),
and shallow eclipses only 2.3\% deep.  The two binaries revolve around
their common center of mass in a highly elongated orbit ($e = 0.85$)
with a period of 16.67\;yr.  Radial velocities are measured for all
components from our quadruple-lined spectra, and are combined with the
light curves and with measurements of times of minimum light for the
2.4 day binary to solve for the elements of the inner and outer orbits
simultaneously. The line-of-sight inclination angles of the three
orbits are similar, suggesting they may be close to coplanar. The
available observations appear to indicate that the 6 day binary
experiences significant retrograde apsidal motion in the amount of
about 60 degrees per century. We derive absolute masses for the four
stars good to better than 1.5\%, along with radii with formal errors
of 1.1\% and 3.5\% for the 2.4 day binary and $\sim$9\% for the 6 day
binary. A comparison of these and other physical properties with
current stellar evolution models gives excellent agreement for a
metallicity of ${\rm [Fe/H]} = -0.15$ and an age of 360\;Myr.
\end{abstract}

\keywords{
binaries: eclipsing ---
stars: evolution ---
stars: fundamental parameters ---
stars: individual (\vstar) ---
techniques: photometric ---
techniques: radial velocities
}

\section{Introduction}
\label{sec:introduction}

The photometric variability of V482\;Persei (alternate designations
BD+47~961, TYC\;3332-314-1; $V = 10.25$, $P = 2.44$\,d) was discovered
photographically by \cite{Hoffmeister:1966} at the Sonneberg
Observatory on the basis of a single instance of a drop in
brightness. The orbital period of 2.44 days was determined later by
\cite{Harvig:1981}, also photographically. The first photoelectric
light curves ($BV$) were published by \cite{Agerer:1991}, along with
several times of minimum light. Continued recording of the times of
eclipse by many authors eventually led to the discovery of the
light-travel time effect \citep{Wolf:2004}, implying the presence of a
third object in the system with a very eccentric ($e \approx 0.82$)
and long-period orbit ($P \approx 16.8$\;yr). Similar parameters for
the third body were reported by \cite{Ogloza:2012}. \cite{Popper:1996}
remarked on an apparent discrepancy between the spectral type implied
by the \cite{Agerer:1991} observations and the weakness of the sodium
D lines. He reported a type of F2. However, the most commonly seen
classification of the star in the literature is A0
\citep[e.g.,][]{Heckmann:1975}, although other sources list the object
as A2 \citep{Luo:2016} or F6 \citep{Pickles:2010}.  More recently
\cite{Basturk:2015} published the first determination of the absolute
properties of the \vstar\ components based on new $BV\!RI$ light
curves and spectroscopic observations.

We placed \vstar\ on our own photometric and spectroscopic observing
program in 2001, also with the goal of deriving accurate physical
properties for the stars. These observations reveal that the object is
in reality a hierarchical quadruple system. Our spectra show four sets
of lines corresponding to the components of two binaries, with the
brighter one having the reported period of 2.4 days and the other, a
period of 6 days. Furthermore, this second binary is also eclipsing
(although the eclipses are very shallow), and both systems orbit a
common center of mass with the 16-year period inferred earlier from
the light-travel time effect. Such quadruple, doubly-eclipsing systems
are relatively rare, though several have been discovered in recent
years based on the high-precision and nearly uninterrupted
observations collected between 2009 and 2013 by NASA's
\kepler\ spacecraft, as well as from ground-based surveys \citep[see,
  e.g.,][]{Pawlak:2013, Koo:2014, Lohr:2015}.

Because \vstar\ has a more complicated nature than it was thought to
have at the time of the analysis by \cite{Basturk:2015}, and because
of the limited spectroscopic material these authors had at their
disposal that did not allow them to resolve the four components, the
properties they derived for the stars in the 2.4 day binary are
incorrect. The motivation for this paper is thus to perform a complete
and independent analysis of our observations with the new knowledge
about the configuration of the system, to determine the physical
properties of all four stars, and to compare them against stellar
evolution models.

We begin in Section\;\ref{sec:observations} by describing our
spectroscopic and photometric observations, as well as the available
times of minimum light for the 2.4 day binary.  Our analysis of these
data is presented in Section\;\ref{sec:analysis}, where we solve for
the orbits of the inner binaries and the outer orbit simultaneously.
The physical properties we determine for the four stars are reported
in Section\;\ref{sec:properties}, and a comparison with stellar
evolution models is found in Section\;\ref{sec:models}. We conclude
with a discussion of the results in Section\;\ref{sec:discussion}.

\section{Observations}
\label{sec:observations}

\subsection{Differential photometry}
\label{sec:photometry}

Differential photometry of \vstar\ was obtained by measuring images
collected with two different robotic telescopes: the URSA WebScope at
the University of Arkansas at Fayetteville, AR \citep{Lacy:2005}, and
the NFO WebScope near Silver City, NM \citep{Grauer:2008}. The URSA
Webscope consists of a 10-inch Meade LX\;200\;SCT with an SBIG\;ST8 CCD
camera, housed in a Technical Innovations RoboDome on top of Kimpel
Hall on campus. The NFO WebScope is a modified Group 128 24-inch
Cassegrain telescope with a CCD camera in a roll-off enclosure.  All
observations were made through a Bessel $V$ filter consisting of
2.0\;mm of GG495 and 3.0\;mm of BG39. Observations were made between
2001 December and 2016 January, and are presented in
Table\;\ref{tab:photursa} and Table\;\ref{tab:photnfo}. Two comparison
stars were measured near the variable star (which has $V = 10.25$, SpT
A0): TYC\;3332-0388-1 ($V = 10.22$, SpT A5) and TYC\;3332-0146-1 ($V =
11.33$). A total of 13,000 frames of \vstar\ were gathered with the
URSA telescope, and 14,072 with the NFO WebScope. All images were
measured with an application ({\tt Measure}) written by author Lacy.
The standard deviations of the differences in magnitudes between the
two comparison stars were 0.012\;mag for the URSA measurements and
0.015\;mag for those from the NFO.

\begin{deluxetable}{cc}
\tablewidth{0pc}
\tablecaption{Differential $V$-band observations of \vstar\ from the URSA
  WebScope\label{tab:photursa}}
\tablehead{
\colhead{HJD-2,400,000} &
\colhead{$\Delta V$ (mag)}
}
\startdata
52250.75606  & 0.019 \\
52250.75697  & 0.023 \\
52250.75789  & 0.029 \\
52250.75879  & 0.020 \\
52250.75971  & 0.031 
\enddata
\tablecomments{This table is available in its entirety in
  machine-readable form.}
\end{deluxetable}

\begin{deluxetable}{cc}
\tablewidth{0pc}
\tablecaption{Differential $V$-band observations of \vstar\ from the NFO
  WebScope\label{tab:photnfo}}
\tablehead{
\colhead{HJD-2,400,000} &
\colhead{$\Delta V$ (mag)}
}
\startdata
53405.79764  & 0.555 \\
53405.80017  & 0.547 \\
53405.80271  & 0.546 \\
53405.80519  & 0.542 \\
53405.80773  & 0.538 
\enddata
\tablecomments{This table is available in its entirety in
  machine-readable form.}
\end{deluxetable}

\subsection{Spectroscopy}
\label{sec:spectroscopy}

\vstar\ was monitored spectroscopically with two different
instruments.  We observed it between 2009 November and 2017 February
at the Harvard-Smithsonian Center for Astrophysics (CfA) with the
Tillinghast Reflector Echelle Spectrograph
\citep[TRES;][]{Szentgyorgyi:2007, Furesz:2008}, a fiber-fed,
bench-mounted instrument on the 1.5\;m Tillinghast reflector at the
Fred L.\ Whipple Observatory (Mount Hopkins, AZ). The wavelength
coverage is approximately 3900--9100\;\AA\ in 51 orders, with a
resolving power of $R \approx 44,000$. For the radial-velocity
measurements described below we used a single order centered on the
\ion{Mg}{1}\,b triplet at 5188\;\AA\ that yields the best results.  A
total of 46 spectra were gathered with typical signal-to-noise ratios
between 30 and 100 per resolution element of 6.8\;\kms. Wavelength
calibrations were based on exposures of a Thorium-Argon lamp taken
before an after each science frame, and radial-velocity standards were
observed each night although they were not used because of the high
stability of the spectrograph ($\sim$20\;\ms, much better than
required for this work).  Reductions were performed with a dedicated
pipeline.

From 2011 November through 2017 April we additionally acquired 37
useful spectra of \vstar\ with the Tennessee State University 2\;m
Automatic Spectroscopic Telescope (AST) and a fiber-fed echelle
spectrograph \citep{Eaton:2007} at Fairborn Observatory in southeast
Arizona. The detector for these observations was a Fairchild 486 CCD,
having a $4096 \times 4096$ array of 15\;$\mu$m pixels. While the
spectrograms have 48 orders ranging from 3800--8260\;\AA, we have used
only the orders that cover the wavelength region from 4920--7100\;\AA.
Because of the faintness of \vstar\ and the moderate rotation of its
components, we made our observations with a fiber that produced a
spectral resolution of 0.4\;\AA, corresponding to a resolving power of
15,000 at 6000\;\AA.  Our spectra have typical signal-to-noise ratios
of 30--40 at this wavelength.  More information about the AST facility
can be found in the paper of \citet{Fekel:2013}.

\begin{figure}
\epsscale{1.0}
\plotone{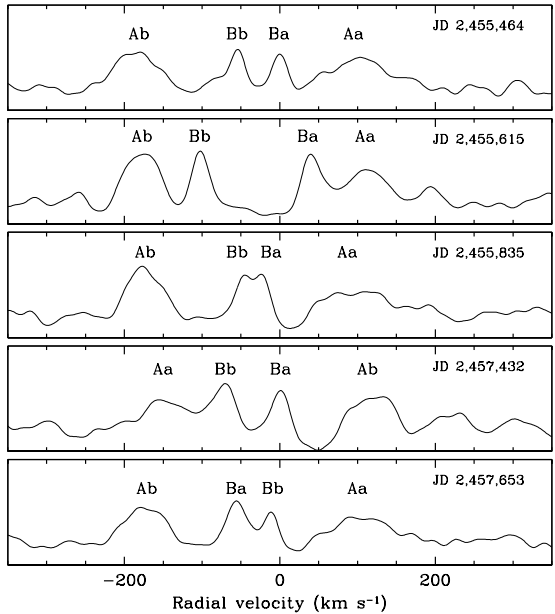} 
\figcaption[]{Examples of one-dimensional cross-correlation
  functions for \vstar\ showing peaks corresponding to the lines of
  the four components, as labeled. The Julian dates are shown in each
  panel.\label{fig:CCFs}}
\end{figure}

Radial-velocity determinations from the TRES spectra proceeded as
follows. Based on the expectation that we would see two sets of
relatively broad lines corresponding to the components of the 2.4 day
binary (binary ``A''), initial determinations of the radial velocities
were made with the two-dimensional cross-correlation technique TODCOR
\citep{Zucker:1994}.  It soon became clear that there were two
additional sets of lines that were much sharper, and did not phase up
with the ephemeris for the 2.4 day binary (see Figure~\ref{fig:CCFs}).
They were eventually found to correspond to the primary and secondary
of a 6 day binary (binary ``B''). Final velocities for the four stars
were then measured with an extension of TODCOR to four dimensions
\citep[QUADCOR;][]{Torres:2007}. Templates (one for each star) were
taken from a large library of synthetic spectra based on PHOENIX model
atmospheres \citep{Husser:2013}. The two main parameters of these
templates, the effective temperature ($T_{\rm eff}$) and rotational
velocity ($v \sin i$ when seen in projection), were determined by
running extensive grids of cross-correlations over wide ranges seeking
the best match to our spectra as measured by the average correlation
coefficient. For an analogous methodology applied to the case of only
two stars, see \cite{Torres:2002}. In this way we determined optimal
temperatures of 10,600\;K and 9600\;K for the components of the 2.4
day binary, referred to in the following as stars Aa (the more massive
one) and Ab. Estimated uncertainties are 200\;K. For each star in the
6 day binary (Ba and Bb, with Ba being marginally more massive; see
Section\;\ref{sec:properties}) we obtained 7600\;K and larger
uncertainties of 300\;K due to their faintness. These temperatures
correspond to spectral types of approximately B9 and A0 for the 2.4
day binary, and A6 for the stars in the 6 day binary
\citep{Gray:1992}. The $v \sin i$ values of stars Aa and Ab were
determined from this procedure to be 60\;\kms\ and 40\;\kms, with
uncertainties of 5\;\kms, and for Ba and Bb we measured $12 \pm
2$\;\kms.  Surface gravities $\log g$ were held at values of 4.0 for
stars Aa and Ab, and 4.5 for Ba and Bb, close to the final values from
our analysis. Solar metallicity was adopted throughout, and modest
changes in composition ($\pm$0.5\;dex in [Fe/H]) have a negligible
effect on the measurements. The final TRES velocities in the
heliocentric frame are listed in Table~\ref{tab:rvs_cfa} along with
their individual uncertainties, which have been adjusted to match the
scatter from a preliminary joint orbital solution for the quadruple
system that used the velocities of all four stars and the eclipse
timings for binary A (see next section). The uncertainties average
3.3, 2.6, 1.3, and 1.3\;\kms\ for stars Aa, Ab, Ba, and Bb,
respectively.  The complicated nature of the spectra makes the flux
ratios among stars difficult to measure. Our estimates with QUADCOR
yield $\ell_{\rm Ab}/\ell_{\rm Aa} = 0.54 \pm 0.04$, $\ell_{\rm
  Ba}/\ell_{\rm Aa} = 0.091 \pm 0.004$, and $\ell_{\rm Bb}/\ell_{\rm
  Aa} = 0.091 \pm 0.004$, and a flux ratio within the 6 day binary of
$\ell_{\rm Bb}/\ell_{\rm Ba} = 1.01 \pm 0.06$, all at the mean
wavelength of our observations, 5188\;\AA.

\begin{deluxetable*}{lrcrcrcrccc}
\tablewidth{0pc}
\tablecaption{Heliocentric radial velocity measurements of
  \vstar\ from CfA\label{tab:rvs_cfa}}
\tablehead{
\colhead{HJD} &
\colhead{$RV_{\rm Aa}$} &
\colhead{$\sigma_{\rm Aa}$} &
\colhead{$RV_{\rm Ab}$} &
\colhead{$\sigma_{\rm Ab}$} &
\colhead{$RV_{\rm Ba}$} &
\colhead{$\sigma_{\rm Ba}$} &
\colhead{$RV_{\rm Bb}$} &
\colhead{$\sigma_{\rm Bb}$} &
\colhead{Phase} &
\colhead{Phase}
\\
\colhead{(2,400,000+)} &
\colhead{(\kms)} &
\colhead{(\kms)} &
\colhead{(\kms)} &
\colhead{(\kms)} &
\colhead{(\kms)} &
\colhead{(\kms)} &
\colhead{(\kms)} &
\colhead{(\kms)} &
\colhead{Aa+Ab} &
\colhead{Ba+Bb}
}
\startdata
 55143.8088 &    44.13 &   6.03 &$-$162.22 &   4.63 & $-$89.18 &   2.33 &    82.77 &   2.28 &  0.8500 &  0.1521 \\ 
 55144.8750 &$-$153.37 &   7.18 &    99.24 &   5.52 & $-$67.54 &   2.77 &    60.70 &   2.72 &  0.2858 &  0.3297 \\ 
 55171.7254 &$-$167.97 &   5.61 &   100.16 &   4.31 &    66.70 &   2.17 & $-$67.21 &   2.12 &  0.2604 &  0.8032 \\ 
 55192.7260 &    41.56 &   3.63 &$-$176.48 &   2.79 & $-$68.08 &   1.40 &    76.58 &   1.38 &  0.8440 &  0.3020 \\ 
 55199.7654 &    50.39 &   3.63 &$-$200.03 &   2.79 &    24.48 &   1.40 & $-$10.75 &   1.37 &  0.7213 &  0.4748 \\ 
 55227.7623 &$-$155.56 &   3.79 &    70.51 &   2.91 & $-$69.42 &   1.47 &    94.66 &   1.43 &  0.1648 &  0.1391 \\ 
 55464.0061 &    82.97 &   2.83 &$-$180.60 &   2.18 &  $-$1.12 &   1.10 & $-$54.90 &   1.07 &  0.7235 &  0.4997 \\ 
 55486.0085 &    72.73 &   3.67 &$-$172.04 &   2.82 &$-$115.84 &   1.42 &    60.88 &   1.39 &  0.7160 &  0.1658 \\ 
 55527.7138 &    90.56 &   3.71 &$-$172.97 &   2.85 &$-$102.76 &   1.43 &    40.82 &   1.40 &  0.7612 &  0.1149 \\ 
 55615.7548 &    88.54 &   1.72 &$-$176.94 &   1.32 &    36.17 &   0.66 &$-$102.89 &   0.65 &  0.7439 &  0.7846 \\ 
 55647.6127 &    81.97 &   3.42 &$-$172.67 &   2.63 & $-$93.14 &   1.32 &    25.58 &   1.30 &  0.7643 &  0.0929 \\ 
 55835.9381 &    87.20 &   2.38 &$-$172.49 &   1.83 & $-$22.29 &   0.92 & $-$48.34 &   0.90 &  0.7332 &  0.4726 \\ 
 55846.9420 &$-$140.60 &   1.82 &   121.07 &   1.40 &$-$109.52 &   0.70 &    39.55 &   0.69 &  0.2305 &  0.3061 \\ 
 55851.9695 &$-$135.74 &   4.56 &   117.66 &   3.50 &$-$118.82 &   1.76 &    50.38 &   1.73 &  0.2852 &  0.1438 \\ 
 55879.9608 &    93.57 &   5.35 &$-$174.91 &   4.11 &    32.57 &   2.07 &$-$101.47 &   2.02 &  0.7253 &  0.8079 \\ 
 55882.9200 &    18.74 &   2.54 & $-$87.11 &   1.96 &$-$109.37 &   0.98 &    42.15 &   0.96 &  0.9347 &  0.3009 \\ 
 55883.9049 &$-$127.93 &   1.96 &    99.27 &   1.50 & $-$24.71 &   0.76 & $-$45.75 &   0.74 &  0.3372 &  0.4650 \\ 
 55884.9629 &    88.13 &   4.12 &$-$180.62 &   3.16 &    33.21 &   1.59 &$-$107.67 &   1.56 &  0.7696 &  0.6413 \\ 
 55906.7672 &    73.51 &   2.71 &$-$160.43 &   2.09 &$-$120.18 &   1.05 &    51.88 &   1.03 &  0.6811 &  0.2745 \\ 
 55910.8002 &$-$130.16 &   2.88 &   100.96 &   2.21 & $-$15.96 &   1.11 & $-$52.73 &   1.09 &  0.3294 &  0.9465 \\ 
 56197.0099 &$-$135.37 &   2.73 &   116.48 &   2.10 &    32.36 &   1.06 &$-$104.79 &   1.03 &  0.3034 &  0.6363 \\ 
 56266.8268 &    69.76 &   3.80 &$-$149.77 &   2.92 &$-$122.71 &   1.47 &    53.06 &   1.44 &  0.8376 &  0.2696 \\ 
 56608.9494 &    67.64 &   3.68 &$-$150.27 &   2.83 &$-$121.79 &   1.42 &    48.33 &   1.39 &  0.6632 &  0.2760 \\ 
 56672.8211 &    89.04 &   4.64 &$-$172.64 &   3.57 &  $-$3.77 &   1.79 & $-$68.09 &   1.76 &  0.7676 &  0.9187 \\ 
 56699.7443 &    91.74 &   4.28 &$-$173.27 &   3.29 & $-$57.27 &   1.66 & $-$13.60 &   1.62 &  0.7711 &  0.4048 \\ 
 56732.6582 &$-$138.50 &   4.78 &   117.79 &   3.67 &     7.54 &   1.85 & $-$78.41 &   1.81 &  0.2230 &  0.8891 \\ 
 56936.0282 &$-$124.81 &   3.22 &   101.00 &   2.47 &    34.86 &   1.24 &$-$108.11 &   1.22 &  0.3404 &  0.7757 \\ 
 57001.6916 &$-$130.48 &   2.41 &   105.33 &   1.85 &    41.24 &   0.93 &$-$110.90 &   0.91 &  0.1770 &  0.7169 \\ 
 57088.7028 &    87.13 &   6.38 &$-$174.30 &   4.90 &$-$132.59 &   2.46 &    60.05 &   2.41 &  0.7386 &  0.2152 \\ 
 57114.6225 &$-$124.49 &   4.21 &   101.39 &   3.24 &     5.14 &   1.63 & $-$76.77 &   1.59 &  0.3320 &  0.5341 \\ 
 57121.6335 &$-$136.65 &   4.67 &   112.56 &   3.59 &    40.67 &   1.80 &$-$106.58 &   1.77 &  0.1974 &  0.7023 \\ 
 57296.9155 &    69.25 &   2.02 &$-$150.51 &   1.55 &     0.31 &   0.78 & $-$70.49 &   0.76 &  0.8352 &  0.9088 \\ 
 57323.8360 &    67.34 &   2.32 &$-$153.87 &   1.79 & $-$62.43 &   0.90 &  $-$5.64 &   0.88 &  0.8376 &  0.3944 \\ 
 57350.7407 &    68.78 &   2.10 &$-$153.30 &   1.61 &    12.60 &   0.81 & $-$81.75 &   0.80 &  0.8336 &  0.8774 \\ 
 57390.6960 &$-$125.54 &   1.97 &    99.88 &   1.51 &     4.14 &   0.76 & $-$76.14 &   0.74 &  0.1634 &  0.5350 \\ 
 57412.7314 &$-$126.66 &   2.31 &   101.63 &   1.78 &$-$129.52 &   0.89 &    61.22 &   0.87 &  0.1693 &  0.2066 \\ 
 57416.7267 &    80.37 &   2.63 &$-$164.31 &   2.02 &    15.91 &   1.02 & $-$84.42 &   1.00 &  0.8021 &  0.8724 \\ 
 57432.6346 &$-$134.43 &   2.44 &   113.41 &   1.87 &     1.51 &   0.94 & $-$69.86 &   0.92 &  0.3037 &  0.5230 \\ 
 57476.6803 &$-$130.11 &   3.58 &   111.41 &   2.75 &    18.78 &   1.38 & $-$87.66 &   1.35 &  0.3052 &  0.8621 \\ 
 57647.9149 &$-$138.57 &   2.08 &   117.12 &   1.60 & $-$63.92 &   0.81 &  $-$5.83 &   0.79 &  0.2890 &  0.3942 \\ 
 57648.9254 &    84.27 &   2.25 &$-$168.77 &   1.73 &    15.51 &   0.87 & $-$83.43 &   0.85 &  0.7020 &  0.5625 \\ 
 57653.9942 &    88.36 &   2.59 &$-$171.63 &   1.99 & $-$55.67 &   1.00 & $-$12.13 &   0.98 &  0.7736 &  0.4071 \\ 
 57679.8731 &$-$118.45 &   2.27 &    94.34 &   1.74 &    41.04 &   0.88 &$-$111.09 &   0.86 &  0.3503 &  0.7192 \\ 
 57706.7676 &$-$122.75 &   2.51 &    95.61 &   1.93 &$-$128.21 &   0.97 &    61.18 &   0.95 &  0.3421 &  0.2005 \\ 
 57766.5938 &    87.44 &   2.36 &$-$163.53 &   1.82 &$-$120.89 &   0.91 &    55.47 &   0.89 &  0.7932 &  0.1690 \\ 
 57794.7257 &$-$135.40 &   2.10 &   114.60 &   1.61 &    20.54 &   0.81 & $-$89.59 &   0.79 &  0.2907 &  0.8565 
\enddata
\end{deluxetable*}

Our AST spectra of \vstar\ also clearly show four sets of lines, and
so, line blending often occurs.  In addition, the average depth of the
lines is only about 1--2\%, and the lines of the 2.4 day binary have
very significant rotational broadening. These factors contribute to
the difficulty in measuring the radial velocities of the stars by the
procedures applied to these spectra.

\citet{Fekel:2009} presented a general explanation of the velocity
measurement of the Fairborn echelle spectra. For the 6 day binary we
used our solar-type star line list to measure velocities because the
lines of the 2.4 day binary are much less visible, and therefore cause
significantly fewer blending problems.  In addition, the solar line
list has more than four times as many lines as the A star line list,
so using that list improves the precision of the averaged
velocities. To measure velocities of the 2.4 day binary we used our A
star line list, which consists mostly of lines of ionized elements.
With that list features of all four stars are visible, and the average
line depth of the four components is similar.

Our velocities were determined by fitting the individual lines with
rotational broadening functions \citep{LacyFekel:2011}, and we allowed
both the depth and width of the line fits to vary. In the case of
blended features we fit both components of the blend simultaneously. A
few velocities of the 6 day binary components were measured with the A
star line list, and those velocities were found to be consistent with
the ones measured with the solar-type star line list. In the end we
obtained 20 pairs of measurements for the 2.4 day binary and 32 pairs
for the 6 day binary.

Our unpublished measurements of several IAU solar-type velocity
standards show that these Fairborn Observatory velocities have a
zero-point offset of $-0.6$\;\kms\ when compared to the results of
\citet{Scarfe:2010}. Thus, we have added 0.6\;\kms\ to each
velocity. Our useful Fairborn observations and the measured
heliocentric velocities are given in Table\;\ref{tab:rvs_ast}.
Typical uncertainties were estimated to be 5.4, 3.9, 1.3, and
1.6\;\kms\ for stars Aa, Ab, Ba, and Bb, based on the scatter from the
preliminary orbital solution mentioned earlier (see also
Section\;\ref{sec:timings}).

\begin{deluxetable*}{crrrrcc}
\tablewidth{0pc}
\tablecaption{Heliocentric radial velocity measurements of
  \vstar\ from the Fairborn Observatory
\label{tab:rvs_ast}}
\tablehead{
\colhead{HJD} &
\colhead{$RV_{\rm Aa}$} &
\colhead{$RV_{\rm Ab}$} &
\colhead{$RV_{\rm Ba}$} &
\colhead{$RV_{\rm Bb}$} &
\colhead{Phase} &
\colhead{Phase}
\\
\colhead{(2,400,000+)} &
\colhead{(\kms)} &
\colhead{(\kms)} &
\colhead{(\kms)} &
\colhead{(\kms)} &
\colhead{Aa+Ab} &
\colhead{Ba+Bb}
}
\startdata
55893.8851 &  $-$87.6  &     50.6  & $-$109.8 &     41.3  &  0.4161 &  0.1280 \\ 
55926.8766 &     42.2  & $-$108.8  &     34.3 & $-$102.7  &  0.8998 &  0.6252 \\ 
55947.8192 &  \nodata  &  \nodata  & $-$106.5 &     33.5  & \nodata &  0.1148 \\ 
55984.6605 &  \nodata  &  \nodata  & $-$126.7 &     56.4  & \nodata &  0.2535 \\ 
56017.6807 &  \nodata  &  \nodata  &     38.4 & $-$108.3  & \nodata &  0.7555 \\ 
56188.9963 &  \nodata  &  \nodata  & $-$110.9 &     38.8  & \nodata &  0.3011 \\ 
56209.7776 &  \nodata  &  \nodata  &     36.4 & $-$109.1  & \nodata &  0.7638 \\ 
56229.7011 &     67.6  & $-$156.2  &  \nodata &  \nodata  &  0.6643 & \nodata \\ 
56265.6515 & $-$111.5  &     94.7  &  $-$86.6 &     14.0  &  0.3572 &  0.0738 \\ 
56288.9111 &     63.9  & $-$136.7  &  \nodata &  \nodata  &  0.8634 & \nodata \\ 
56328.8122 & $-$124.5  &    100.5  &     27.7 &  $-$98.0  &  0.1710 &  0.5980 \\ 
56353.6641 & $-$131.4  &    107.1  &     39.4 & $-$109.9  &  0.3280 &  0.7389 \\ 
56559.9008 &  \nodata  &  \nodata  & $-$100.9 &     31.2  & \nodata &  0.1033 \\ 
56630.6538 &  \nodata  &  \nodata  &      6.8 &  $-$77.3  & \nodata &  0.8925 \\ 
56649.6758 & $-$142.6  &    119.1  &  $-$78.5 &      8.6  &  0.3081 &  0.0621 \\ 
56667.6360 &     73.2  & $-$145.5  &  $-$72.4 &      4.9  &  0.6484 &  0.0547 \\ 
56686.8530 &  \nodata  &  \nodata  & $-$126.0 &     57.7  & \nodata &  0.2568 \\ 
56702.7574 &  \nodata  &  \nodata  &      1.9 &  $-$70.0  & \nodata &  0.9068 \\ 
56931.8161 &  \nodata  &  \nodata  &  $-$85.9 &     16.0  & \nodata &  0.0739 \\ 
56951.7638 &     95.0  & $-$171.9  &  $-$60.4 &   $-$7.6  &  0.7715 &  0.3977 \\ 
56992.0182 & $-$135.6  &    122.9  & $-$100.8 &     33.6  &  0.2235 &  0.1051 \\ 
57297.7976 & $-$136.8  &    123.4  &  \nodata &  \nodata  &  0.1957 & \nodata \\ 
57359.7054 &  \nodata  &  \nodata  &  $-$75.0 &      5.9  & \nodata &  0.3712 \\ 
57401.8024 &     93.6  & $-$172.9  &  $-$69.9 &   $-$3.1  &  0.7026 &  0.3856 \\ 
57434.7679 & $-$138.5  &    105.1  &     13.7 &  $-$80.7  &  0.1756 &  0.8785 \\ 
57464.7014 &  \nodata  &  \nodata  &     16.6 &  $-$85.2  & \nodata &  0.8662 \\ 
57607.9271 &  \nodata  &  \nodata  &     39.9 & $-$107.6  & \nodata &  0.7312 \\ 
57649.8716 &  $-$79.1  &     50.8  &     37.8 & $-$110.1  &  0.0887 &  0.7202 \\ 
57676.8915 & $-$116.3  &     77.5  & $-$128.1 &     62.9  &  0.1317 &  0.2224 \\ 
57694.9427 &  \nodata  &  \nodata  & $-$129.2 &     61.9  & \nodata &  0.2302 \\ 
57711.9514 &  \nodata  &  \nodata  &  $-$77.4 &     10.5  & \nodata &  0.0642 \\ 
57735.8273 & $-$141.5  &    119.6  &  \nodata &  \nodata  &  0.2189 & \nodata \\ 
57789.7710 & $-$148.6  &    121.2  &  $-$58.2 &  $-$12.6  &  0.2657 &  0.0310 \\ 
57814.7642 &  \nodata  &  \nodata  & $-$130.8 &     61.0  & \nodata &  0.1954 \\ 
57820.7494 &  \nodata  &  \nodata  & $-$126.3 &     58.8  & \nodata &  0.1927 \\ 
57833.7036 & $-$140.3  &    115.2  &  $-$87.7 &     20.1  &  0.2210 &  0.3512 \\ 
57849.6362 &     87.2  & $-$173.2  &  \nodata &  \nodata  &  0.7327 & \nodata   
\enddata
\tablecomments{Uncertainties for stars Aa, Ab, Ba, and Bb are 5.4,
  3.9, 1.3, and 1.6\;\kms, respectively.}
\end{deluxetable*}

Rotational broadening fits of lines in our spectra that have the
highest signal-to-noise ratios result in $v \sin i$ values of $59 \pm
5$\;\kms\ and $39 \pm 3$\;\kms\ for the primary and secondary of the
2.4 day binary, respectively. For the 6 day binary components we
determine $v \sin i$ values of $11 \pm 2$\;\kms\ and $13 \pm
2$\;\kms\ for stars Ba and Bb.

From the best Fairborn spectra, the average line equivalent width
ratio of the components in the 6 day orbit is $0.97 \pm 0.07$, which
should be indicative of the true light ratio $\ell_{\rm Bb}/\ell_{\rm
  Ba}$ at all wavelengths as their effective temperatures are
essentially the same. This measurement is consistent with our estimate
from the CfA spectra.

\subsection{Times of minimum light}
\label{sec:timings}

Numerous times of minimum light have been recorded for the 2.4 day
binary since its discovery. The few photographic estimates reported by
\cite{Hoffmeister:1966} and \cite{Harvig:1981} are too poor to be
useful for the present work. The other, more recent determinations are
collected in Table~\ref{tab:timings} along with their reported
uncertainties, where available.  A total of 78 correspond to eclipses
of star Aa, and 36 to those of Ab. They span 27.5 years, or about 1.6
cycles of the outer 16-year orbit between the A and B binaries.

Experience indicates that published uncertainties for this type of
observation are not always accurate, and are often underestimated.  To
test this we carried out a solution for the outer orbit that used the
times of minimum along with the radial velocities described earlier,
modeling the third-body effect on the timings with the classical
formalism by \cite{Irwin:1952, Irwin:1959}. Based on the residuals
from this fit we established that the primary and secondary timing
errors require scale factors of about 2.8 and 4.5 in order to obtain
reduced $\chi^2$ values near unity. Similarly, for measurements with
no published errors we found average uncertainties of 0.0027\;days and
0.0010\;days to be suitable for the primary and secondary timings,
respectively. We adjusted the published errors accordingly, and
adopted them for our analysis in Section\;\ref{sec:solution}. The same
procedure was used to adjust the uncertainties for the radial
velocities, as mentioned before, arriving at the values reported in
Tables\;\ref{tab:rvs_cfa} and \ref{tab:rvs_ast}.

\begin{deluxetable}{ll@{}c@{}cc@{}c@{}}
\tablecaption{Times of minimum light for \vstar.\label{tab:timings}}
\tablehead{
\colhead{HJD} &
\colhead{$\sigma$} &
\colhead{} &
\colhead{$(O-C)$} &
\colhead{} &
\colhead{} \\
\colhead{(2,400,000$+$)} &
\colhead{(days)} &
\colhead{Type} &
\colhead{(days)} &
\colhead{Year} &
\colhead{Source}
}
\startdata
  47565.3737  & \nodata  &   2  & $-$0.00286 &  1989.104  &   1 \\
  47823.5048  & \nodata  &   1  & $-$0.00420 &  1989.811  &   1 \\
  47840.636   & \nodata  &   1  & $-$0.00026 &  1989.858  &   1 \\
  47850.4210  & \nodata  &   1  & $-$0.00225 &  1989.885  &   1 \\
  47943.4012  & \nodata  &   1  & $+$0.00150 &  1990.139  &   1 
\enddata
\tablecomments{Measurement errors ($\sigma$) are listed as published,
  when available. Uncertainties for the timings with no published
  errors are assumed to be 0.0027\;days for primary minima and
  0.0010\;days for secondary minima (see Section\;\ref{sec:solution}).
  ``Type'' is 1 for a primary eclipse, 2 for a secondary eclipse.
  $O\!-\!C$ residuals are computed from the combined fit described in
  Section~\ref{sec:analysis}.  Sources for the times of minimum light
  are:
(1) \cite{Agerer:1991};
(2) \cite{Hubscher:1991};
(3) \cite{Hubscher:1992};
(4) \cite{Hubscher:1993};
(5) \cite{Hubscher:1994};
(6) \cite{Agerer:1995};
(7) \cite{Agerer:1996};
(8) \cite{Agerer:1997};
(9) \cite{Agerer:1998};
(10) \cite{Agerer:1999};
(11) \cite{Paschke:2017};
(12) \cite{Lacy:2002};
(13) \cite{Lacy:2003};
(14) \cite{Agerer:2003};
(15) \cite{Kotkova:2006};
(16) \cite{Zejda:2004};
(17) \cite{Lacy:2004};
(18) \cite{Lacy:2006};
(19) \cite{Brat:2007};
(20) \cite{Lacy:2007};
(21) \cite{Hubscher:2007};
(22) \cite{Lacy:2009};
(23) \cite{Yilmaz:2009};
(24) \cite{Hubscher:2009a};
(25) \cite{Hubscher:2009b};
(26) \cite{Diethelm:2009};
(27) \cite{Lacy:2011};
(28) \cite{Diethelm:2011a};
(29) \cite{Diethelm:2011b};
(30) \cite{Hubscher:2012};
(31) \cite{Liakos:2011};
(32) \cite{Lacy:2012};
(33) \cite{Lacy:2013};
(34) \cite{Diethelm:2013};
(35) \cite{Hubscher:2015};
(36) \cite{Jurysek:2017};
(37) \cite{Zasche:2017}.
This table is available in its entirety in machine-readable form.
}
\end{deluxetable}

\begin{figure}
\epsscale{1.10}
\plotone{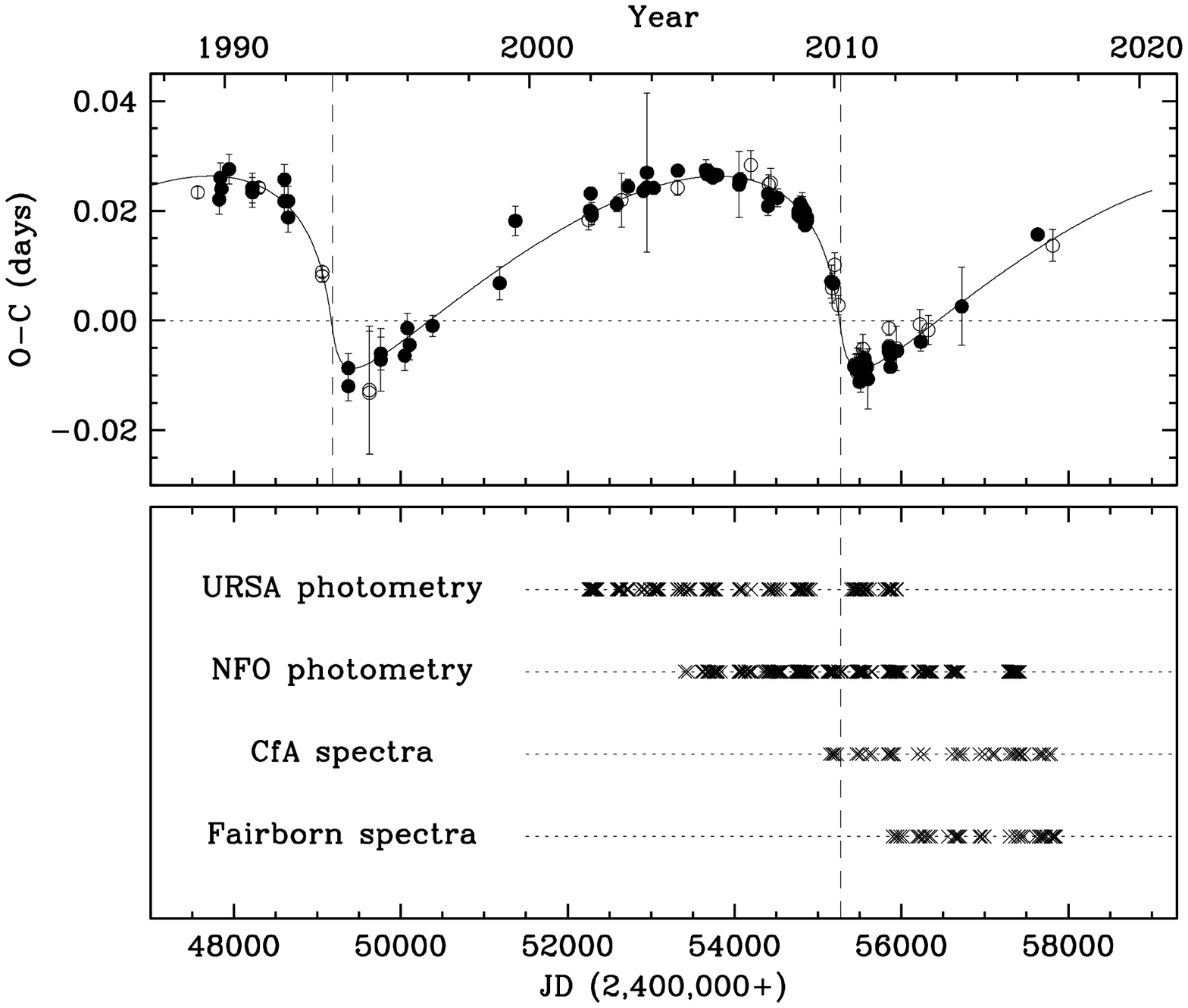}

\figcaption{{\it Top:} Times of minimum light from
  Table\;\ref{tab:timings} with our best-fit linear ephemeris
  subtracted out, to yield observed minus computed ($O-C$) residuals
  showing the light-travel time effect. Filled and open symbols
  represent primary and secondary minima, respectively, and the solid
  curve is our best fit model for the light-travel effect described in
  Section\;\ref{sec:solution}. The vertical dashed lines indicate
  times of periastron passage in the outer orbit.  {\it Bottom:} Time
  history of our photometric and spectroscopic observations for
  comparison with the eclipse timings.  \label{fig:timings}}

\end{figure}

The top panel of Figure\;\ref{fig:timings} shows all timing
measurements after subtracting the linear ephemeris from our best-fit
global orbital solution described below. They display the obvious
light-travel time effect (LTTE) first reported by \cite{Wolf:2004},
and provide a strong constraint on the elements of the outer
orbit. Also shown in the figure is the time history of our other
observations.  The early CfA spectra were gathered fortuitously near
periastron passage.

\section{Analysis}
\label{sec:analysis}

The configuration of the quadruple \vstar\ system is hierarchical,
with the outer period being 1000 times longer than the longest of the
inner periods (see below). For the purposes of this work we will
regard this architecture to be sufficiently well represented by three
non-interacting Keplerian orbits. The different types of observations
available constrain the parameters of the three orbits in different
ways, and are quite complementary. The approach in this paper is
therefore to combine them all into a single solution to make optimal
use of the information.  There is in fact some redundancy such that
some of the elements can be obtained in more than one way, as
described below, and this allows one to reduce the number of
variables. It is also possible to constrain others properties of the
system not usually accessible in non-eclipsing binaries.

\subsection{Constraints on orbital elements}
\label{sec:constraints}

In our initial analysis our model for the system included only the
eclipses of the 2.4 day binary, guided by previous work and the
appearance of the light curves. The usual spectroscopic elements for
this binary are $P_{\rm A}$, $K_{\rm Aa}$, $K_{\rm Ab}$, $e_{\rm A}$,
$\omega_{\rm A}$, and $T_{\rm A}$, where the longitude of periastron
$\omega$ refers to the primary (star Aa) and $T_{\rm A}$ is a
reference time of primary eclipse. The inclination angle is $i_{\rm
  A}$.  There was no obvious evidence in the raw photometry of
eclipses of the 6 day binary, despite the fact that preliminary
spectroscopic orbital solutions suggested large and nearly equal
minimum masses for its components close to what was expected from the
temperatures of the stars, indicative of a high inclination angle.
Because we measure velocities for all four stars, the outer ``binary''
(A\;+\;B) is effectively double-lined, and its velocity semiamplitudes
$K_{\rm A}$ and $K_{\rm B}$ can be determined directly.  Under these
circumstances it is possible to infer the inclination angle of the 6
day binary from elements of the three orbits as
\begin{equation}
\label{eq:sin3iB}
\sin^3 i_{\rm B} = \frac
{P_{\rm B} (1-e_{\rm B}^2)^{3/2} (K_{\rm Ba}+K_{\rm Bb})^3}
{P_{\rm A} (1-e_{\rm A}^2)^{3/2} (K_{\rm Aa}+K_{\rm Ab})^3}
\frac{K_{\rm B}}{K_{\rm A}}\sin^3 i_{\rm A}~,
\end{equation}
and thus to obtain the absolute masses of the four stars.

It is also possible to infer the inclination angle of the outer orbit,
which is given in terms of other known elements by
\begin{equation}
\label{eq:sin3iAB}
\sin^3 i_{\rm AB} = \frac
{P_{\rm AB} (1-e_{\rm AB}^2)^{3/2} (K_{\rm A}+K_{\rm B})^2 K_{\rm B}}
{P_{\rm A} (1-e_{\rm A}^2)^{3/2} (K_{\rm Aa}+K_{\rm Ab})^3}
\sin^3 i_{\rm A}~,
\end{equation}
where subscripts ``AB'' refer to the outer orbit.

Closer examination subsequently revealed very shallow eclipses of the
6 day binary precisely at the phases expected from the spectroscopic
orbit (see below), allowing a direct measurement of $i_{\rm B}$ and
enabling the absolute masses of its components to be determined
another way (without recourse to the outer orbit). This redundancy
yields a relation between the semiamplitudes of the outer orbit (from
Eq.\;\ref{eq:sin3iB}) as
\begin{equation}
\label{eq:KB}
K_{\rm B} = \frac
{P_{\rm A} (1-e_{\rm A}^2)^{3/2} (K_{\rm Aa}+K_{\rm Ab})^3}
{P_{\rm B} (1-e_{\rm B}^2)^{3/2} (K_{\rm Ba}+K_{\rm Bb})^3}
\frac{\sin^3 i_{\rm B}}{\sin^3 i_{\rm A}}K_{\rm A}~,
\end{equation}
and can be used to eliminate either $K_{\rm A}$ or $K_{\rm B}$ as
adjustable variables. Our spectroscopic observations alone give us
relatively weak constraints on these semiamplitudes because the radial
velocities cover only a fraction of the outer orbit (though they do
partially sample periastron passage). On the other hand, the times of
minimum light of Section\;\ref{sec:timings} span more than one cycle
of the outer orbit, and help to pin down $K_{\rm A}$, with the net
effect that this quantity is better constrained by the observations
than $K_{\rm B}$.  Consequently, we have chosen to use
Eq.\;\ref{eq:KB} to eliminate $K_{\rm B}$, retaining only $K_{\rm A}$
as a free parameter.

\subsection{Solution}
\label{sec:solution}

Both inner binaries are well detached. We modeled their light curves
using the Nelson-Davis-Etzel formalism \citep{Popper:1981,
  Etzel:1981}, as implemented in the widely used EBOP code, which is
adequate for systems such as these with nearly spherical stars. In
order to allow the flexibility to incorporate various constraints
described below, and to combine all observations together solving
simultaneously for all parameters, we made use of a version of EBOP
due to \cite{Irwin:2011} that is especially useful within the
framework of the Markov Chain Monte Carlo methodology we apply
here.\footnote{\url https://github.com/mdwarfgeek/eb~.} The relative
weighting of the different data sets relied on the uncertainties
established for each type of observation, as described earlier, and we
verified that modest changes in those uncertainties did not affect the
results significantly.

In addition to the ephemeris ($P_{\rm A}$ and the time of primary
minimum $T_{\rm A}$), the light-curve elements for the 2.4 day binary
are the $V$-band central surface brightness ratio between the
secondary and the primary $J_{\rm A}$, the sum of the relative radii
$r_{\rm Aa} + r_{\rm Ab}$, the radius ratio $k_{\rm A} \equiv r_{\rm
  Ab}/r_{\rm Aa}$, the cosine of the inclination angle $\cos i_{\rm
  A}$, the eccentricity parameters $\sqrt{e_{\rm A}}\cos\omega_{\rm
  A}$ and $\sqrt{e_{\rm A}}\sin\omega_{\rm A}$, and a third-light
parameter $L_3$ to account for the dilution effect produced by the
flux from the other binary, where $L_3$ is the fractional light
contribution relative to the total. Similar adjustable light-curve
parameters were considered for the 6 day binary, once we discovered it
is also eclipsing. Both sets of parameters were solved for
simultaneously, with the third-light parameter for binary B being
simply $1-L_3$.  We also solved for separate out-of-eclipse magnitude
levels for URSA and NFO, $m_{\rm URSA}$ and $m_{\rm NFO}$, and allowed
for separate scale factors applied to the estimated internal
photometric errors from these two telescopes, $f_{\rm URSA}$ and
$f_{\rm NFO}$, which were set initially to values of 0.008\;mag and
0.012\;mag, respectively, from preliminary fits.  Test solutions
indicated negligible eccentricity in the 2.4 day orbit, so for the
final fit we considered it to be circular.

Limb darkening was represented with the linear law, as experiments
with a two-parameter quadratic law gave no improvement. The $V$ band
coefficients for the four components were taken from the tables of
\cite{Claret:2011} in accordance with the stellar properties reported
earlier. They are 0.451, 0.494, 0.597, and 0.597 for stars Aa, Ab, Ba,
and Bb, respectively. Gravity darkening coefficients were calculated
as described by \cite{Torres:2017}, and were set to 0.672, 0.727,
0.886, and 0.886.

Additional spectroscopic parameters of the fit are the center-of-mass
velocity of the quadruple system $\gamma$, the velocity semiamplitudes
of the inner orbits $K_{\rm Aa}$, $K_{\rm Ab}$, $K_{\rm Ba}$, and
$K_{\rm Bb}$, the period and reference epoch of periastron passage of
the outer orbit $P_{\rm AB}$ and $T_{\rm AB}$, the eccentricity
parameters $\sqrt{e_{\rm AB}}\cos\omega_{\rm AB}$ and $\sqrt{e_{\rm
    AB}}\cos\omega_{\rm AB}$ (where $\omega_{\rm AB}$ corresponds to
the ``primary'' in the outer orbit, i.e., the A binary), and the
velocity semiamplitude $K_{\rm A}$ mentioned earlier, tracing the
motion of the center of mass of the A binary. Additionally we allowed
for a possible difference $\Delta RV$ in the velocity zero points of
our CfA and Fairborn observations. Initially we considered also
possible offsets between the primary and secondary velocity zero
points within each inner binary, which may result, e.g., from template
mismatch in the CfA determinations. We found these offsets to be
insignificant in early tests, and therefore dropped them for the final
solutions.

The periodic variations in the times of minimum light of the A binary
were modeled as mentioned earlier with the third-body formalism of
\cite{Irwin:1952, Irwin:1959}. For all practical purposes these
measurements may be assumed to correspond to times of conjunction, as
the difference is negligible in our case. The LTTE in the outer orbit
was fully accounted for in the treatment of the radial velocity motion
in the inner orbits, and for the light curve solutions. This was done
by adjusting the individual times of observation at each step of the
iterations based on the current values of the outer elements.  All
reference epochs from our solution ($T_{\rm A}$, $T_{\rm B}$, $T_{\rm
  AB}$) are in the frame of the center of mass of the quadruple
system, and were defined to be near the average epoch of all
observations, to minimize correlations.

Our method of solution for \vstar\ used the {\tt
  emcee\/}\footnote{\url http://dan.iel.fm/emcee~.} code of
\cite{Foreman-Mackey:2013}, which is a Python implementation of the
affine-invariant Markov Chain Monte Carlo (MCMC) ensemble sampler
proposed by \cite{Goodman:2010}. We used 300 walkers, and uniform
priors over suitable ranges for most elements. Initial solutions
showed that the radius ratio $k_{\rm B}$ in the 6 day binary was
poorly constrained from photometry alone, and converged to
unrealistically low values under 0.5 for two stars that are in fact
very similar in mass and temperature. A similar problem occurred with
the central surface brightness ratio $J_{\rm B}$. This is not
surprising given the very shallow eclipses caused by heavy dilution
from the much brighter 2.4 day binary, the presence of instrumental
errors in one of our photometric data sets (NFO) that may be
distorting the eclipse shapes and depths (see below), and the partial
nature of the eclipses of two similar stars \citep[see,
  e.g.,][]{Andersen:1991}.

To overcome this difficulty we made use of the measured spectroscopic
light ratio between stars Ba and Bb ($\ell_{\rm Bb}/\ell_{\rm Ba} =
0.99 \pm 0.05$, the weighted average of our two determinations from
Section\;\ref{sec:spectroscopy}), which is strongly correlated with
the radius ratio ($\ell_{\rm Bb}/\ell_{\rm Ba} \propto k_{\rm B}^2$),
and applied it as a Gaussian prior on the light ratio to constrain the
fit. And because the two stars appear to have essentially identical
temperatures, we also used a Gaussian prior of $1.00 \pm 0.02$ on the
central surface brightness ratio $J_{\rm B}$.  An additional
constraint imposed on our solutions was that $\sin^3 i_{\rm AB}$, as
given by Eq.\;\ref{eq:sin3iAB}, be strictly less than unity (with
$K_{\rm B}$ computed from Eq.\;\ref{eq:KB}).  Convergence was checked
by examining the chains visually, and verifying that the Gelman-Rubin
statistic \citep{Gelman:1992, Brooks:1997} was smaller than 1.05 for
all adjustable parameters.

Early solutions resulted in a satisfactory fit to most of the
observations except for the velocities of the 6 day binary, which
showed an obvious pattern of phase-dependent residuals far in excess
of the estimated uncertainties. It was eventually found that this
could be eliminated by allowing for apsidal motion in this slightly
eccentric orbit. The addition of $\dot\omega_{\rm B}$ as a free
parameter to our fit did indeed yield a highly significant value of
about 60 degrees per century, but with a sign that indicated
precession in the direction {\it opposite} to the orbital motion. We
discuss this further below.  With allowance for apsidal motion, the
orbital period of the B binary we solved for is strictly the sidereal
period, which in this case is longer than the anomalistic period for
the reason indicated.  The fitted value of $\omega_{\rm B}$ is for the
reference epoch of primary eclipse, $T_{\rm B}$.

\begin{deluxetable}{lc}
\tablecaption{Combined orbital solution for \vstar.\label{tab:results1}}
\tablehead{
\colhead{~~~~~~~~~~~~Parameter~~~~~~~~~~~~} &
\colhead{Value and uncertainty}
}
\startdata
$P_{\rm A}$ (days)\dotfill & $2.44675265 \pm 0.00000027$ \\
$T_{\rm A}$ (HJD)\tablenotemark{a}\dotfill & $2,\!454,\!848.10959 \pm 0.00032$ \\
$J_{\rm A}$\dotfill &  $0.778 \pm 0.047$ \\
$r_{\rm Aa}+r_{\rm Ab}$\dotfill & $0.3560 \pm 0.0053$ \\
$k_{\rm A} \equiv r_{\rm Ab}/r_{\rm Aa}$\dotfill & $0.644 \pm 0.021$ \\
$\cos i_{\rm A}$\dotfill & $0.119 \pm 0.018$ \\
$L_3$\tablenotemark{b}\dotfill & $0.145 \pm 0.049$ \\
$f_{\rm URSA}$\dotfill & $0.9929 \pm 0.0069$ \\
$f_{\rm NFO}$\dotfill & $1.0596 \pm 0.0073$ \\
$m_{\rm URSA}$ (mag)\dotfill & $0.0312 \pm 0.0012$ \\
$m_{\rm NFO}$ (mag)\dotfill & $0.3269 \pm 0.0010$ \\
$P_{\rm B}$ (days)\dotfill & $6.001749 \pm 0.000023$ \\
$T_{\rm B}$ (HJD)\tablenotemark{a}\dotfill & $2,\!454,\!848.8326 \pm 0.0071$ \\
$J_{\rm B}$\dotfill & $0.973 \pm 0.043$ \\
$r_{\rm Ba}+r_{\rm Bb}$\dotfill & $0.136 \pm 0.012$ \\
$k_{\rm B} \equiv r_{\rm Bb}/r_{\rm Ba}$\dotfill & $1.009 \pm 0.044$ \\
$\sqrt{e_{\rm B}}\cos\omega_{\rm B}$\dotfill & $-0.3317 \pm 0.0046$ \\
$\sqrt{e_{\rm B}}\sin\omega_{\rm B}$\dotfill & $-0.0164 \pm 0.0095$ \\
$\dot\omega_{\rm B}$ ($10^{-5}$ rad day$^{-1}$)\dotfill & $-2.89 \pm 0.62$ \\
$\cos i_{\rm B}$\dotfill & $0.081 \pm 0.018$ \\
$\gamma$ (\kms)\dotfill & $-30.03 \pm 0.14$ \\
$\Delta RV$ (\kms)\tablenotemark{c}\dotfill & $-0.21 \pm 0.22$ \\
$K_{\rm Aa}$ (\kms)\dotfill & $115.60 \pm 0.86$ \\
$K_{\rm Ab}$ (\kms)\dotfill & $147.22 \pm 0.52$ \\
$K_{\rm Ba}$ (\kms)\dotfill & $85.02 \pm 0.34$ \\
$K_{\rm Bb}$ (\kms)\dotfill & $85.72 \pm 0.35$ \\
$P_{\rm AB}$ (days)\dotfill & $6089 \pm 29$ \\
$T_{\rm AB}$ (HJD)\tablenotemark{a}\dotfill & $2,\!455,\!271.5 \pm 5.2$ \\
$K_{\rm A}$ (\kms)\dotfill & $17.20 \pm 0.75$ \\
$\sqrt{e_{\rm AB}}\cos\omega_{\rm AB}$\dotfill & $-0.8631 \pm 0.0066$ \\
$\sqrt{e_{\rm AB}}\sin\omega_{\rm AB}$\dotfill & $-0.3291 \pm 0.0097$ 
\enddata

\tablecomments{The values reported correspond to the mode from the
  MCMC posterior distributions. The uncertainties come from the
  residual permutation procedure described in the text.}

\tablenotetext{a}{$T_{\rm A}$ and $T_{\rm B}$ are the reference times
  of primary eclipse in the 2.4 day and 6 day binaries (eclipse of
  stars Aa and Ba), and $T_{\rm AB}$ is the reference time of
  periastron passage in the outer orbit.}

\tablenotetext{b}{Fraction of the light contributed by stars Ba+Bb.}

\tablenotetext{c}{Zero-point difference between the CfA and Fairborn
  velocity frames, in the sense Fairborn minus CfA.}

\end{deluxetable}

\begin{deluxetable}{lc}
\tablecaption{Derived properties for \vstar.\label{tab:results2}}
\tablehead{
\colhead{~~~~~~~~~~~~Parameter~~~~~~~~~~~~} &
\colhead{Value and uncertainty}
}
\startdata
$r_{\rm Aa}$\dotfill & $0.2166 \pm 0.0021$ \\
$r_{\rm Ab}$\dotfill & $0.1393 \pm 0.0048$ \\
$r_{\rm Ba}$\dotfill & $0.0681 \pm 0.0058$ \\
$r_{\rm Bb}$\dotfill & $0.0686 \pm 0.0061$ \\
$K_{\rm LTTE,A}$ (min)\tablenotemark{a}\dotfill & $25.24 \pm 0.25$ \\

$P_{\rm B,anom}$ (days)\tablenotemark{b}\dotfill & $6.001583 \pm 0.000015$ \\
$T_{\rm peri,B}$ (HJD)\tablenotemark{c}\dotfill & $2,\!454,\!850.171 \pm 0.029$ \\
$\dot\omega_{\rm B}$ (deg century$^{-1}$)\dotfill & $-60 \pm 13$ \\
$U_{\rm B}$ (yr)\tablenotemark{d}\dotfill & $590 \pm 210$ \\
$e_{\rm B}$\dotfill & $0.1103 \pm 0.0031$  \\
$\omega_{\rm B}$ (deg)\tablenotemark{e}\dotfill & $182.9 \pm 1.6$ \\
$\phi_{\rm B}$\tablenotemark{f}\dotfill & $0.4300 \pm 0.0020$ \\
$P_{\rm AB}$ (yr)\dotfill & $16.672 \pm 0.079$ \\
$K_{\rm B}$ (\kms)\tablenotemark{g}\dotfill & $26.4 \pm 1.2$ \\
$e_{\rm AB}$\dotfill & $0.8533 \pm 0.0084$ \\
$\omega_{\rm AB}$ (deg)\tablenotemark{e}\dotfill & $200.90 \pm 0.67$ \\
$\sin^3 i_{\rm AB}$\tablenotemark{h}\dotfill & $0.958 \pm 0.056$ \\
$i_{\rm AB}$ (deg)\dotfill & $79.6 \pm 6.4$ \\
$i_{\rm A}$ (deg)\dotfill & $83.2 \pm 1.0$ \\
$i_{\rm B}$ (deg)\dotfill & $85.3 \pm 1.1$ \\
$\ell_{\rm Ab}/\ell_{\rm Aa}$\dotfill & $0.324 \pm 0.029$ \\
$\ell_{\rm Ba}/\ell_{\rm Aa}$\dotfill & $0.112 \pm 0.042$  \\
$\ell_{\rm Bb}/\ell_{\rm Aa}$\dotfill & $0.113 \pm 0.042$ \\
$\ell_{\rm Bb}/\ell_{\rm Ba}$\dotfill & $1.000 \pm 0.019$ \\
$\ell_{\rm B}/\ell_{\rm A}$\dotfill & $0.169 \pm 0.065$ \\
$q_{\rm A} \equiv M_{\rm Ab}/M_{\rm Aa}$\dotfill & $0.7853 \pm 0.0065$ \\
$q_{\rm B} \equiv M_{\rm Bb}/M_{\rm Ba}$\dotfill & $0.9918 \pm 0.0050$ \\
$q_{\rm AB} \equiv M_{\rm B}/M_{\rm A}$\dotfill & $0.652 \pm 0.010$ \\
$a_{\rm A}$ ($R_{\sun}$)\dotfill & $12.800 \pm 0.050$ \\
$a_{\rm B}$ ($R_{\sun}$)\dotfill & $20.199 \pm 0.067$ \\
$a_{\rm AB}$ (au)\dotfill & $12.926 \pm 0.054$ 
\enddata

\tablecomments{The values reported correspond to the mode from the
  MCMC posterior distributions. The uncertainties come from the
  residual permutation procedure described in the text. The physical
  constants used to calculate the semimajor axes conform to IAU
  recommendations from 2015 Resolution B3 \citep[see][]{Prsa:2016}.}

\tablenotetext{a}{Semi-amplitude of the light-travel time effect on the
  eclipse timings of the 2.4 day binary, caused by motion in the orbit
  around the center of mass with the 6 day binary.}

\tablenotetext{b}{Anomalistic period in the 6 day binary.}

\tablenotetext{c}{Time of periastron passage for the 6 day binary.}

\tablenotetext{d}{Apsidal period in the 6 day binary.}

\tablenotetext{e}{Following the spectroscopic convention, the angle
  $\omega_{\rm B}$ corresponds to star Ba, and $\omega_{\rm AB}$ to
  binary A in the wide orbit.}

\tablenotetext{f}{Phase of secondary eclipse in the 6 day binary orbit.}

\tablenotetext{g}{Computed with Eq.\;\ref{eq:KB}.}

\tablenotetext{h}{Computed with Eq.\;\ref{eq:sin3iAB}.}

\end{deluxetable}

The resulting 31 parameters from our MCMC solution are presented in
Table\;\ref{tab:results1}, and other properties derived from the
fitted parameters are listed in
Table\;\ref{tab:results2}.\footnote{For quantities that are
  combinations of others, our choice to report the mode of all
  posterior distributions can result in small, unavoidable differences
  between the mode of the derived quantity and the results one would
  compute directly from the modal values of the independent variables
  (such as $r_{\rm Aa}$ or $r_{\rm Ab}$ from the radius sum and
  $k_{\rm A}$).} The formal uncertainties returned by the procedure
were found to be too small because it does not account for
time-correlated noise in our observations (``red'' noise), which is
significant particularly in the differential NFO photometry, as
discussed below. To address this concern we carried out a residual
permutation (``prayer bead'') exercise in which we shifted the
residuals from our original fit by an arbitrary number of time indices
(for all data sets), added them back into the model curves at each
time of observation (with wrap-around over each data set), and then
performed the MCMC adjustment again on the synthetic data sets.  This
preserves the pattern of the correlated noise.  We also perturbed both
the limb darkening and the gravity darkening coefficients by adding
Gaussian noise with $\sigma = 0.1$.  We repeated this operation 100
times, and adopted the scatter from the distribution of results for
each adjusted and derived parameter as the final uncertainty. We
consider these error estimates to be more realistic: they are
typically 2--10 times larger than the internal errors.

\begin{figure*}
\epsscale{1.15}
\plotone{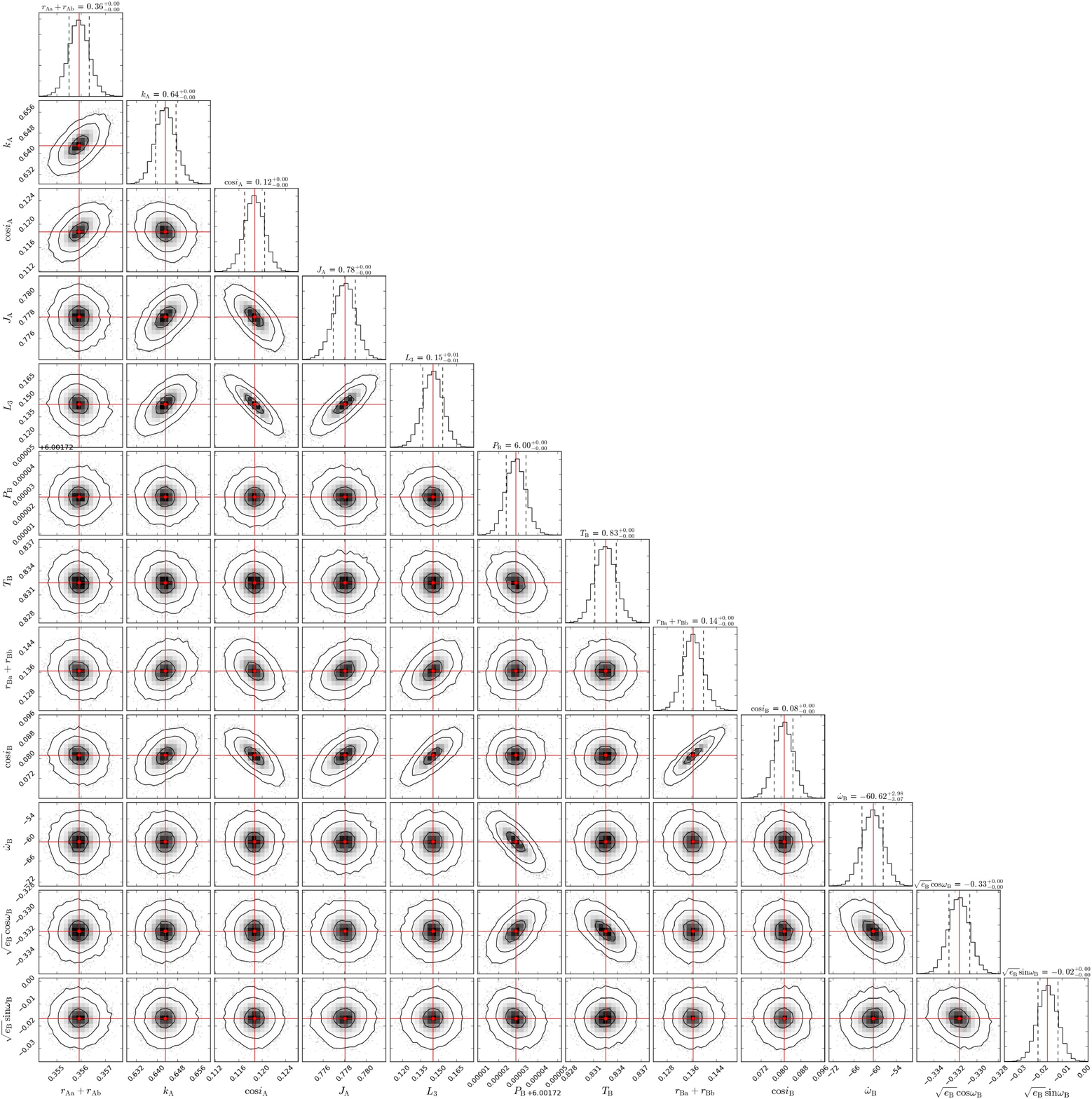}

\figcaption{``Corner plot'' \citep{Foreman-Mackey:2016}\footnote{\url
    https://github.com/dfm/corner.py~.} for \vstar\ illustrating the
  correlations among a selection of the fitted parameters of our
  solution.  Contour levels correspond to 1, 2, and 3$\sigma$, and the
  histograms on the diagonal represent the posterior distribution for
  each parameter, with the mode and internal 68\% confidence levels
  indicated. More realistic errors are discussed in the
  text.\label{fig:corner}}

\end{figure*}

As may be expected from the complexity of the solution several of the
fitted parameters are quite strongly correlated (for example,
\{$r_{\rm Aa}+r_{\rm Ab}$, $k_{\rm A}$\}, \{$J_{\rm A}$, $\cos i_{\rm
  A}$, $L_3$\}, \{$r_{\rm Ba}+r_{\rm Bb}$, $\cos i_{\rm B}$\},
\{$P_{\rm B}$, $\dot\omega_{\rm B}$\}, \{$J_{\rm A}$, $\cos i_{\rm
  A}$, $\cos i_{\rm B}$, $L_3$\}, etc.).  We used the chains from our
Monte Carlo analysis to illustrate this in Figure\;\ref{fig:corner},
for some of the variables with the strongest correlations. Others not
shown that are also correlated include \{$P_{\rm AB}$, $K_{\rm A}$,
$\sqrt{e_{\rm AB}}\cos \omega_{\rm AB}$, $\sqrt{e_{\rm AB}}\sin
\omega_{\rm AB}$\} and \{$\gamma$, $\Delta RV$\}.

Despite the use of the MCMC method that is designed to explore the
high-dimensional parameter space more thoroughly than traditional
least-squares techniques, the uniqueness of a solution with as many
adjustable parameters as we have is generally difficult to prove,
particularly in the presence of significant correlations among some of
the variables, as shown above. We investigated this by repeating our
solution using different sets of initial values for the parameters,
which causes MCMC to sweep parameter space in a different way each
time. We found that in all cases the results were consistent with
those we report.

\begin{figure}
\epsscale{1.10}
\plotone{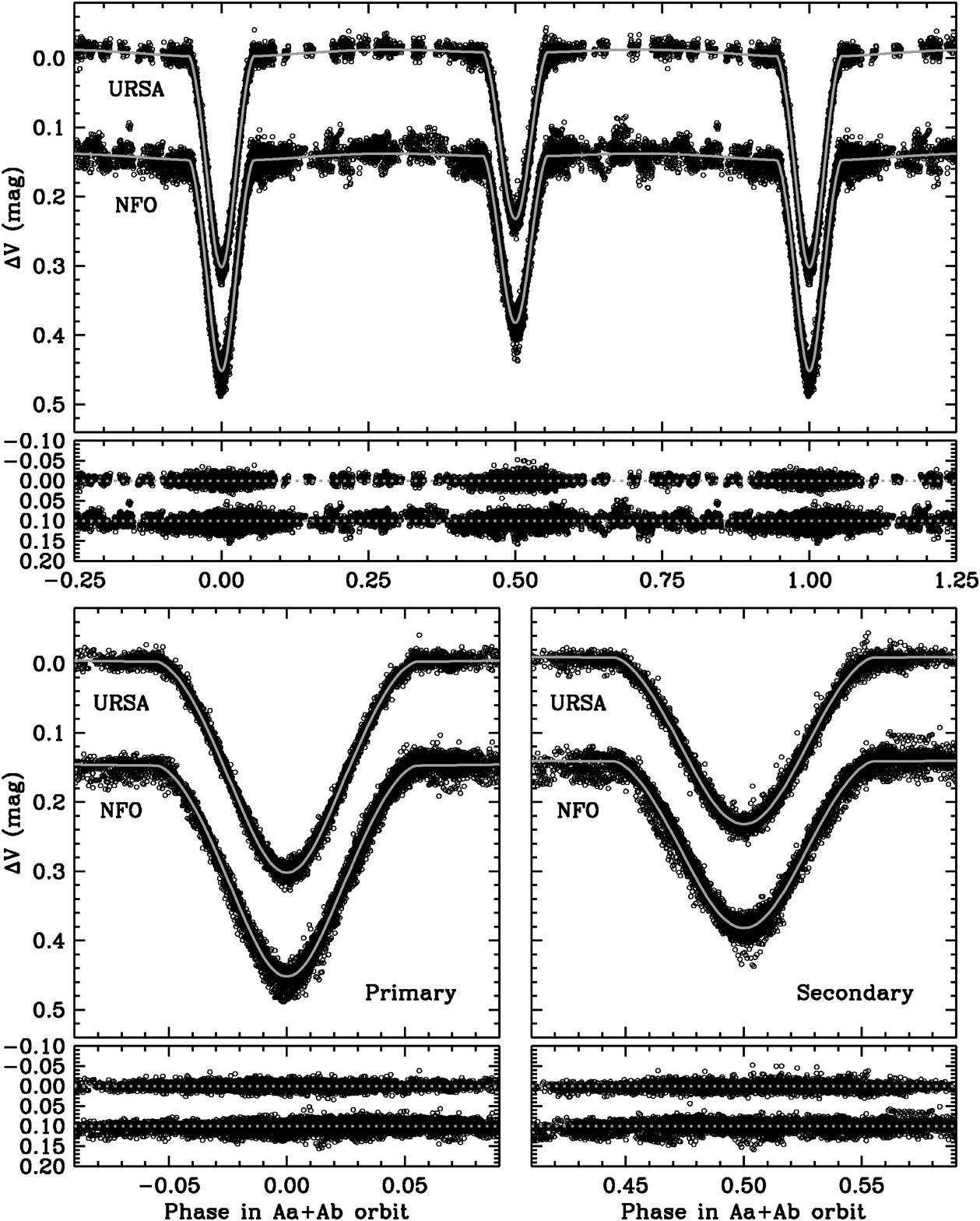}
\figcaption{Differential $V$-band photometry of \vstar\ from the URSA
  and NFO telescopes, along with our best-fit eclipse model for the
  2.4 day binary. Enlargements of the primary and secondary minima are
  also shown. Residuals are displayed beneath each panel. The NFO data
  and residuals are offset vertically for clarity, and the eclipses of
  the 6 day binary have been removed from the
  observations.\label{fig:eclA}}
\end{figure}

\begin{figure}
\epsscale{1.10}
\plotone{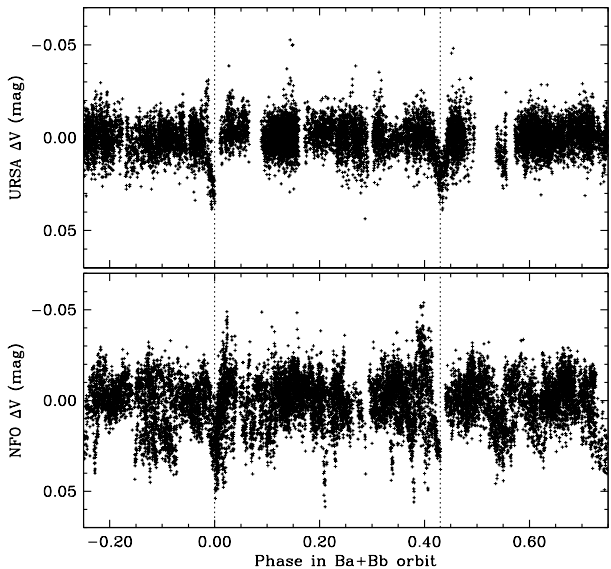}

\figcaption{Residuals from the URSA and NFO observations after
  subtracting the variations in the 2.4 day binary based on our global
  best-fit model.\label{fig:ursanfoB}}

\end{figure}

\begin{figure}
\centering
\includegraphics[angle=-90,width=0.46\textwidth]{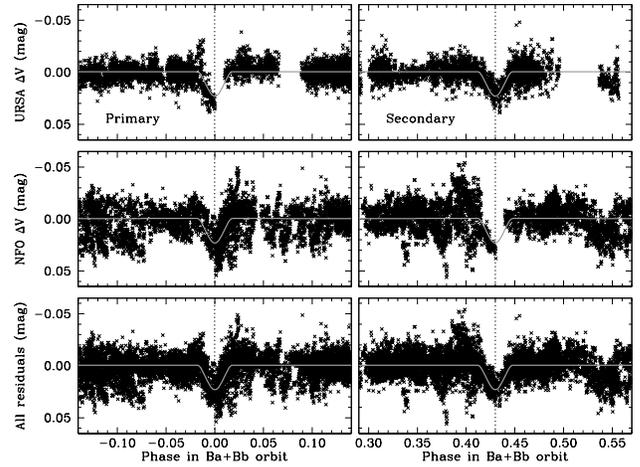}

\figcaption{Enlargement of Figure\;\ref{fig:ursanfoB} around the
  eclipse regions, along with our best-fit model for the 6 day
  binary. The observations are combined in the lower
  panels.\label{fig:eclB}}

\end{figure}

Figure\;\ref{fig:eclA} shows our differential photometry compared with
the best-fit model for the 2.4 day binary. The eclipses in the 6 day
binary have been removed from the data. Conversely, subtracting the
variations in the 2.4 day binary from the original data gives the
residuals seen in Figure\;\ref{fig:ursanfoB}, displayed separately for
the URSA and NFO telescopes. An enlargement of the eclipse regions is
shown in Figure\;\ref{fig:eclB} (top and middle panels) along with the
best-fit eclipse model for the 6 day binary.  Despite the only partial
coverage near phase 0.0, the URSA observations show clear evidence of
dips in the light curve at the precise locations where eclipses could
occur in this binary, according to the spectroscopic orbit (vertical
dotted lines). On the other hand, the evidence for eclipses in the NFO
data is marginal. As discussed by \cite{Lacy:2008}, NFO observations
are known to suffer from small but significant offsets from night to
night that appear to be up to a few hundredths of a magnitude for
\vstar. They are due to a combination of centering errors and
responsivity variations across the field of view.  The URSA
observations are much less affected. The NFO systematics are clearly
visible in Figures\;\ref{fig:ursanfoB} and \ref{fig:eclB}, and are in
fact comparable to the size of the eclipses in the 6 day binary, which
are measured to be only 0.023\;mag deep.  As a result, the evidence
for eclipses in binary B is not particularly compelling from the NFO
data alone, especially with their lack of coverage near the secondary
minimum. These features might have been missed entirely were it not
for the independent URSA data and the critical information from
spectroscopy. The bottom panels of Figure\;\ref{fig:eclB} show the
URSA and NFO observations combined, which provide full coverage of
both eclipses. The primary and secondary minima are equally deep.

Our detection of statistically significant apsidal precession in the 6
day binary in a direction opposite to that of the orbital motion
should be taken with caution.  In principle such an effect may well
arise from the hierarchical configuration of the \vstar\ system, with
a highly eccentric outer orbit that could lead to dynamical
interactions between the inner binaries (see
Section\;\ref{sec:discussion}).  However, the presence of systematic
errors in at least one of our photometric data sets that are of the
same order as the depths of the shallow eclipses of the 6 day binary,
combined with the small number of nights in which the eclipses were
observed, have the potential to bias the measurement of
$\dot\omega_{\rm B}$ although it is unclear by how much.
Unfortunately the data in hand are less than optimal for an
independent check as both the URSA and NFO observations each miss one
of the eclipses. A solution without the NFO data still indicated a
significant negative apsidal motion.  Additional independent
observations are highly desirable to confirm this result. In any case,
we note that the absolute dimensions of the stars (masses, radii)
reported below are unaffected by $\dot\omega_{\rm B}$.

\begin{figure}
\epsscale{1.10}
\plotone{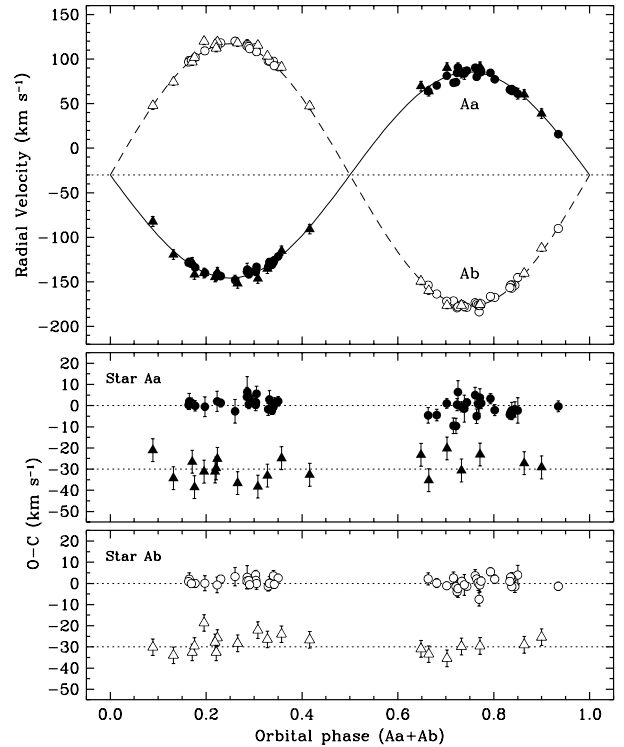}
\figcaption{Radial velocity observations of \vstar\ in the 2.4 day
  binary with our best-fit model. Motion in the outer orbit has been
  subtracted. The dotted line represents the center-of-mass velocity
  of the quadruple system. Residuals are shown at the bottom,
  separately for the CfA and Fairborn measurements (circles and
  triangles, respectively).\label{fig:orbitA}.}
\end{figure}

\begin{figure}
\epsscale{1.10}
\plotone{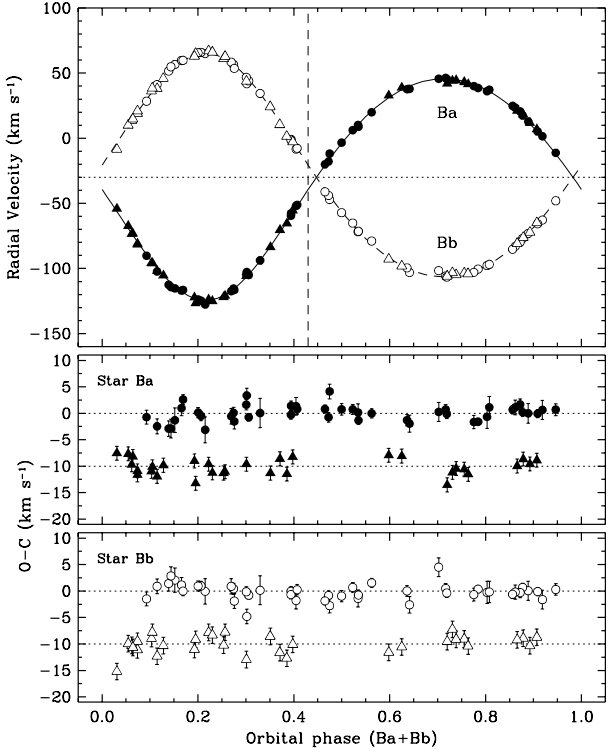}
\figcaption{Same as Figure\;\ref{fig:orbitA} for the 6 day binary. The
  vertical dashed line indicates the phase of secondary eclipse at
  $\phi_{\rm B} = 0.4300 \pm 0.0020$, according to our
  fit.\label{fig:orbitB}}
\end{figure}

\begin{figure}
\epsscale{1.14}
\plotone{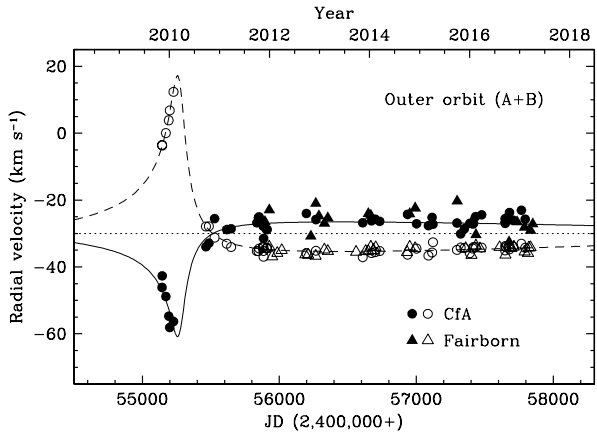}
\figcaption{Radial velocity motion of the centers of mass of the 2.4
  and 6 day binaries in the outer 16\;yr orbit.  The individual
  measurements for each star have been corrected for the motion in the
  corresponding inner binary, and then weight-averaged together for
  display purposes and represented by a single symbol. Filled symbols
  correspond to Aa+Ab, and open ones to Ba+Bb. The solid curves
  represent our best fit, and the dashed line marks the center-of-mass
  velocity of the quadruple system.\label{fig:outer}.}
\end{figure}

The spectroscopic observations of the 2.4 day and 6 day binary
components are shown in Figures\;\ref{fig:orbitA} and
\ref{fig:orbitB}, along with the best-fit models and residuals. In
each case we have subtracted the motion in the 16\;yr outer orbit for
display purposes.  The spectroscopic coverage of the outer orbit is
illustrated in Figure\;\ref{fig:outer}. The symbols represent
instantaneous measurements of the RV of the center of mass of each
binary, calculated by taking the weighted average of the individual
primary and secondary velocity residuals after removing the motion in
the inner orbits. The early CfA observations are seen to have been
taken near the important periastron phase. Finally, the fit to the
times of minimum light for the 2.4 day binary may be seen in
Figure\;\ref{fig:timings}, presented earlier.

\section{Physical properties}
\label{sec:properties}

In Table\;\ref{tab:dimensions} we summarize the physical properties of
the four stars in \vstar\ derived from the parameters of our orbital
solution in the previous section. Stars Ba and Bb have nearly
identical masses, and their radii are indistinguishable within the
errors.

\begin{deluxetable}{lcc}
\tablewidth{0pt}
\tablecaption{Physical properties of the \vstar\ components.\label{tab:dimensions}}
\tablehead{
\colhead{~~~~~~~~~~Parameter~~~~~~~~~~} &
\colhead{Primary} &
\colhead{Secondary}
}
\startdata
\multicolumn{3}{c}{Binary Aa+Ab} \\
\noalign{\hrule\vskip 2pt}
$M$ ($M_{\sun}$)\dotfill                           & $2.634 \pm 0.029$    &  $2.068 \pm 0.030$ \\ 
$R$ ($R_{\sun}$)\dotfill                           & $2.774 \pm 0.031$    &  $1.784 \pm 0.062$  \\ 
$\log g$ (cgs)\dotfill                             & $3.9727 \pm 0.0083$  &  $4.251 \pm 0.030$  \\ 
$T_{\rm eff}$ (K)\dotfill                          & 10600~$\pm$~200\phn\phn      &  9600~$\pm$~200\phn \\ 
$L/L_{\sun}$\dotfill                               & $87.5 \pm 6.9$\phn     &  $24.4 \pm 2.6$\phn  \\ 
$BC_{\rm V}$ (mag)\tablenotemark{a}\dotfill        & $-$0.38~$\pm$~0.11\phs       &    $-$0.17~$\pm$~0.11\phs  \\ 
$M_{\rm bol}$ (mag)\tablenotemark{b}\dotfill       & $-0.123 \pm 0.085$\phs   &  $1.27 \pm 0.12$ \\ 
$M_V$ (mag)\dotfill                                & 0.26~$\pm$~0.11              &  1.43~$\pm$~0.13  \\ 
$E(B-V)$ (mag)\dotfill                             & \multicolumn{2}{c}{0.36~$\pm$~0.06} \\ 
$m-M$ (mag)\dotfill                                & \multicolumn{2}{c}{9.36~$\pm$~0.22} \\ 
Distance (pc)\tablenotemark{c}\dotfill             & \multicolumn{2}{c}{$746 \pm 75$\phn} \\ 
Parallax (mas)\dotfill                             & \multicolumn{2}{c}{$1.34 \pm 0.14$} \\ 
$v_{\rm sync} \sin i$ (\kms)\tablenotemark{d}\dotfill  & $56.96 \pm 0.65$\phn  &  $36.63 \pm 0.65$\phn \\ 
$v \sin i$ (\kms)\tablenotemark{e}\dotfill         &   60~$\pm$~5\phn             &    40~$\pm$~5\phn \\
$v \sin i$ (\kms)\tablenotemark{e}\dotfill         &   59~$\pm$~5\phn             &    39~$\pm$~3\phn  \\
\noalign{\vskip 2pt\hrule\vskip 2pt}
\multicolumn{3}{c}{Binary Ba+Bb} \\
\noalign{\hrule\vskip 2pt}
$M$ ($M_{\sun}$)\dotfill                           & $1.540 \pm 0.016$  &  $1.528 \pm 0.016$ \\ 
$R$ ($R_{\sun}$)\dotfill                           & $1.37 \pm 0.12$    &  $1.39 \pm 0.13$  \\ 
$\log g$ (cgs)\dotfill                             & $4.350 \pm 0.073$  &  $4.338 \pm 0.077$  \\ 
$T_{\rm eff}$ (K)\dotfill                          & 7600~$\pm$~300\phn           &  7600~$\pm$~300\phn \\ 
$L/L_{\sun}$\dotfill                               & $5.6 \pm 1.3$       &  $5.8 \pm 1.4$  \\ 
$BC_{\rm V}$ (mag)\tablenotemark{a}\dotfill        & +0.03~$\pm$~0.10\phs         &    +0.03~$\pm$~0.10\phs  \\ 
$M_{\rm bol}$ (mag)\tablenotemark{b}\dotfill       & $2.85 \pm 0.25$       &  $2.82 \pm 0.26$ \\ 
$M_V$ (mag)\dotfill                                & $2.82 \pm 0.27$       &  $2.79 \pm 0.28$  \\ 
$E(B-V)$ (mag)\dotfill                             & \multicolumn{2}{c}{0.34~$\pm$~0.08} \\ 
$m-M$ (mag)\dotfill                                & \multicolumn{2}{c}{9.24~$\pm$~0.50} \\ 
Distance (pc)\tablenotemark{c}\dotfill             & \multicolumn{2}{c}{$700 \pm 170$} \\ 
Parallax (mas)\dotfill                             & \multicolumn{2}{c}{$1.42 \pm 0.32$} \\ 
$v_{\rm sync} \sin i$ (\kms)\tablenotemark{d}\dotfill  & $11.5 \pm 1.0$\phn  &  $11.7 \pm 1.0$\phn \\ 
$v_{\rm psync} \sin i$ (\kms)\tablenotemark{d}\dotfill & $12.4 \pm 1.1$\phn  &  $12.5 \pm 1.2$\phn \\ 
$v \sin i$ (\kms)\tablenotemark{e}\dotfill         &   12~$\pm$~2\phn             &    12~$\pm$~2\phn \\
$v \sin i$ (\kms)\tablenotemark{e}\dotfill         &   11~$\pm$~2\phn             &    13~$\pm$~2\phn 
\enddata

\tablenotetext{a}{Bolometric corrections from \cite{Flower:1996}, with
  a contribution of 0.10 mag added in quadrature to the uncertainty
  from the temperatures.}

\tablenotetext{b}{Uses $M_{\rm bol}^{\sun} = 4.732$ for consistency
  with the adopted table of bolometric
  corrections \citep[see][]{Torres:2010}.}

\tablenotetext{c}{Relies on the luminosities, the apparent
  magnitude of \vstar\ out of eclipse \citep[$V = 10.250 \pm
    0.032$;][]{Zacharias:2015}, and bolometric corrections.}

\tablenotetext{d}{Projected synchronous and pseudo-synchronous
  rotational velocities.}

\tablenotetext{e}{Measured values from our CfA and Fairborn spectra.}

\end{deluxetable}

The distance to each binary was computed independently relying on
bolometric corrections from \cite{Flower:1996}, the out-of-eclipse
magnitude of the system \citep[$V = 10.250 \pm
  0.032$;][]{Zacharias:2015}, our third-light estimate from
Table\;\ref{tab:results1}, and (distance-dependent) reddening
estimates from \cite{Green:2015} determined by iterations to reach
convergence. The reddening values for the two binaries are in good
agreement, as are the derived distances. Similar results for the
distance were obtained using the radiative flux scale and absolute $V$
magnitude calibration of \cite{Popper:1980}. We note that the adopted
reddening values from \cite{Green:2015} are different (larger) than
most estimates from other sources \citep{Hakkila:1997, Schlegel:1998,
  Drimmel:2003, Amores:2005}, which is possibly explained by
uncertainties due to the low Galactic latitude of the object
($-2.4\arcdeg$). The parallax estimates for the two binaries are
formally two to four times more precise than the entry for \vstar\ in
the {\it Gaia}/DR1 catalog \citep{Lindegren:2016}, $\pi_{\rm DR1} =
0.56 \pm 0.57$\;mas.

The measured $v \sin i$ values for the stars in the 2.4 day binary are
consistent with estimates for synchronous rotation ($v_{\rm sync}\sin
i$), while those for the 6 day binary cannot distinguish between
synchronous and pseudo-synchronous rotation \citep{Hut:1981}.

The masses of all four components are formally determined to better
than 1.5\%, and the radii of stars Aa and Ab to 1.1\% and 3.5\%,
respectively. The radii of stars Ba and Bb are considerably worse
($\sim$9\%), largely on account of systematic errors in the
observations (``red noise'').  Because of the complicated nature of
the orbital solution we consider the radii to be less robust than the
masses, and systematic errors that are difficult to quantify may
contribute further to the uncertainties we have reported. While some
external information was already used above to impose priors on our
fit and strengthen the determination of otherwise poorly constrained
quantities related to the 6 day binary ($k_{\rm B}$, $J_{\rm B}$), the
independent constraints we have available to check the accuracy of
some of the derived properties for the 2.4 day binary are relatively
weak.  For example, if we make the reasonable assumption that the spin
axes of the stars in the 2.4 day binary are parallel to the orbital
axis, and that their rotations are synchronized with the orbital
motion, as seems to be the case (see above), then the ratio of our
measured $v \sin i$ values for the components should be equal to the
radius ratio $k_{\rm A}$. The projected rotational velocities from our
Fairborn spectra yield $v_{\rm Ab}\sin i/v_{\rm Aa}\sin i = 0.66 \pm
0.08$, which agrees with the much more precise $k_{\rm A}$ value
listed in Table\;\ref{tab:results1}. The estimate from the CfA spectra
($0.67 \pm 0.10$) is even more uncertain and therefore less useful,
but still agrees.

Our spectroscopic light ratios from the CfA spectra allow further
checks. The $\ell_{\rm Ba}/\ell_{\rm Aa}$ and $\ell_{\rm Bb}/\ell_{\rm
  Aa}$ values, converted from the mean wavelength of 5188\;\AA\ to the
$V$ band\footnote{The conversion was performed using synthetic spectra
  by \cite{Husser:2013} based on PHOENIX model atmospheres for the
  adopted temperatures of the components, and the radius ratios from
  Table\;\ref{tab:results2}.} yield a ratio of about 0.10 that is not
far from the determinations listed in Table\;\ref{tab:results2}
($\sim$0.11).  However, we find a discrepancy in the Ab/Aa ratio. Our
spectroscopic estimate from Section\;\ref{sec:spectroscopy} converted
to the $V$ band is $0.55 \pm 0.04$, which is considerably larger than
measured from the light curve solution ($\sim$0.32). Given that the
spectroscopy and the light curve fit produce consistent results for
the flux ratios between each of the two stars in the 6 day binary and
star Aa, the problem would appear to be with star Ab. Although in
principle an error in our adopted temperature for that star could bias
the spectroscopic light ratio (but is unlikely to affect the radial
velocities), tests suggest the required change in $T_{\rm eff}$ is
much too large. Alternatively, we speculate that a bias in the
spectroscopic light ratio could occur if star Ab were chemically
peculiar (i.e., a metallic-line A star), in which case our synthetic
templates would not be a good match to the real star. A-type stars
with such anomalies are overwhelmingly members of binary systems and
rotate more slowly than A stars in the field.  The measured rotation
rate of star Ab ($\sim$40\;\kms) is in fact slow enough to be in the
range where these chemical anomalies are seen in other binaries.
Confirmation of this hypothesis would require a detailed chemical
analysis.

\section{Comparison with stellar evolution models}
\label{sec:models}

\begin{figure}
\epsscale{1.10}
\plotone{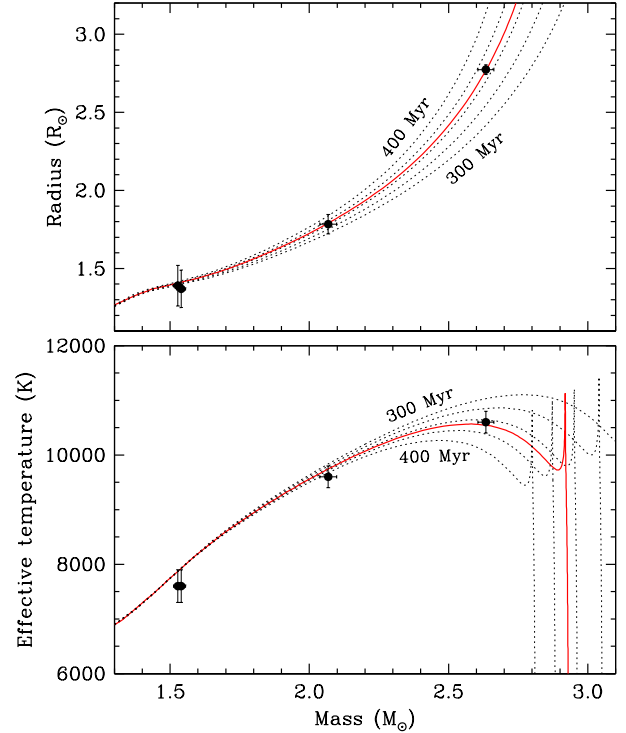}

\figcaption{Measurements for \vstar\ compared against model isochrones
  from the MIST series \citep{Choi:2016} for a metallicity of ${\rm
    [Fe/H]} = -0.15$ and ages of 300--400\;Myr, in steps of 25\;Myr.
  The best match is indicated by the solid curve line, and corresponds
  to an age of 360\;Myr. \label{fig:mist_iso}}

\end{figure}

\begin{figure}
\epsscale{1.10}
\plotone{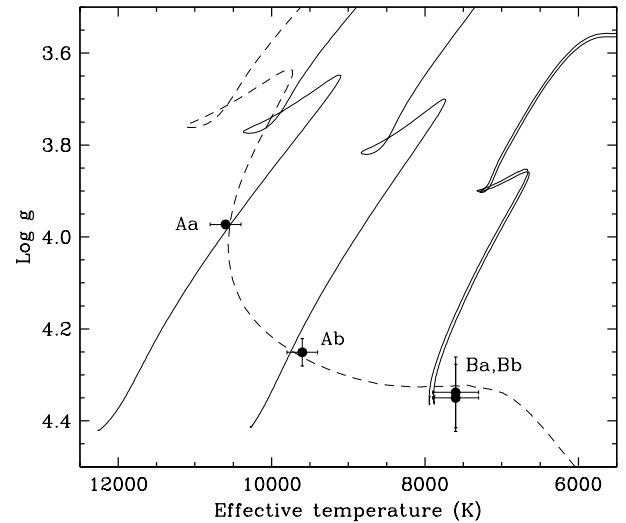}

\figcaption{Evolutionary tracks from the MIST series \citep{Choi:2016}
  for the measured masses of the \vstar\ components in the $T_{\rm
    eff}$--$\log g$ plane. The dashed line represents the best-fit
  isochrone for an age of 360\;Myr and ${\rm [Fe/H]} =
  -0.15$.\label{fig:mist_tracks}}

\end{figure}

Eclipsing binaries with well-determined masses, radii, and
temperatures provide some of strongest tests of stellar evolution
theory available \citep{Torresetal:2010}. The presence of four stars
in \vstar\ sharing a common age and chemical composition offers an
even stronger test.  Figure\;\ref{fig:mist_iso} presents a comparison
of the observations for \vstar\ against model isochrones from the
recent MESA Isochrones and Stellar Tracks (MIST) series by
\cite{Choi:2016}, which is based on the Modules for Experiments in
Stellar Astrophysics package \citep[MESA;][]{Paxton:2011, Paxton:2013,
  Paxton:2015}. The mass-radius diagram in the top panel indicates
excellent agreement between theory and observation for a metallicity
of ${\rm [Fe/H]} = -0.15$ and an age of 360\;Myr constrained mostly by
the properties of stars Aa and Ab. The slightly sub-solar composition
suggested by the models is close enough to the solar value adopted
throughout our analysis that it has a negligible effect on our
measurements of the system.  The agreement with the stellar
temperatures shown in the lower panel of the figure is also good for
Aa and Ab; the Ba and Bb components appear only marginally cooler than
predicted. The state of evolution of each star is seen more clearly in
Figure\;\ref{fig:mist_tracks}, which shows evolutionary tracks from
these models computed for the measured masses, along with the same
best-fit isochrone as above. Stars Ba and Bb are near the zero-age
main-sequence, while Aa is more than halfway through its main sequence
lifetime.

Other models give similar results. For example, a comparison against
isochrones from the Yonsei-Yale series \citep{Yi:2001, Demarque:2004}
yields a good match to the observations for an age of 375\;Myr and a
composition of ${\rm [Fe/H]} = -0.28$. The difference in the best-fit
compositions is simply a consequence of the adoption of different
solar metallicities in these two series of models. MIST adopts the
solar element mixture by \cite{Asplund:2009}, giving a metal content
$Z_{\sun} = 0.0134$, whereas the Yonsei-Yale models adopt the mixture
of \cite{Grevesse:1996}, in which $Z_{\sun} = 0.0179$. The higher
value for the latter models should then result in a [Fe/H] scale that
is $\log(0.0179/0.0134) \approx 0.13$\;dex lower, which is precisely
what we find. The small age difference between the two models is
likely related to differences in their physical ingredients, such as
the treatment of convective core overshooting and the helium abundance.

\section{Discussion and final remarks}
\label{sec:discussion}

Binary stars and systems of higher multiplicity provide valuable
insights into star formation and the role of dynamical and dissipative
processes in shaping the architecture of stellar systems.  Statistical
studies indicate that hierarchical quadruple systems are relatively
rare. \cite{Tokovinin:2014} reported a rate of occurrence among F- and
G-type stars of only 4\%, while \cite{DeRosa:2014} found a smaller
rate of about 2\% among A stars. \vstar\ is remarkable in that we are
able to measure radial velocities of the four components and that both
inner binaries display eclipses, yielding direct measurements of the
masses and radii for all stars at a single age and composition. The
first known example of such a doubly-eclipsing quadruple system is
BV\;Dra + BW\;Dra \citep{Batten:1965, Batten:1986}, a wide visual pair
with a 16\arcsec\ angular separation that enables the two W\;UMa
eclipsing binaries to be studied separately.

Several more doubly-eclipsing quadruple systems have been found that
feature complex light curves, but most have not yet been studied
spectroscopically. Aside from \vstar, we know of only four cases in
which it has been possible to measure radial velocities for all
components from the quadruple-lined spectra to derive their physical
properties: V994\;Her \citep{Lee:2008, Zasche:2013, Zasche:2016},
KIC\;4247791 \citep{Lehmann:2012}, 1SWASP\,J093010.78+533859.5
\citep[possibly a quintuple system;][]{Koo:2014, Lohr:2015}, and
EPIC\;220204960 \citep{Rappaport:2017}. The outer orbit is known in
only one of these cases (V994\;Her), and in \vstar.  Additional
examples of quadruple systems have been found in which one of the
inner pairs eclipses, but not the other. A partial list includes
LO\;Hya \citep[with inner periods of 2.50\;days and 5.97\;days that
  are remarkably similar to those in \vstar; see][]{Fekel:1981,
  Bakos:1985, Docobo:2007}, V379\;Cep \citep{Harmanec:2007},
BD$-$22\;5866 \citep{Shkolnik:2008}, and KIC\;7177553
\citep{Lehmann:2016}.

The \vstar\ system appears dynamically stable. To verify this we used
the 3-body hierarchical stability criterion of \cite{Eggleton:1995}
treating each binary as the perturbing third body for the other.  The
minimum period ratios for stability are $P_{\rm outer}/P_{\rm inner}
\sim 200$, whereas the observed values are about 2500 for binary A and
1000 for binary B.

As reported in Section\;\ref{sec:solution}, one of the intriguing
findings of the present investigation is the fairly large and
apparently significant rate of {\it retrograde} apsidal precession in
the slightly eccentric 6 day binary, in the amount of about
60\;deg\;century$^{-1}$. Based on the measured stellar properties and
theoretical internal structure constants ($\log k_2$ of $-2.41$ and
$-2.40$ for stars Ba and Bb, from \citealt{Claret:2004}), the rate of
apsidal motion one would expect for the 6 day binary is $2.62 \pm
0.16$\;deg\;century$^{-1}$ in the prograde sense, of which 82\%, or
2.15\;deg\;century$^{-1}$, is due to General Relativity. Provided our
measurement of $\dot\omega_{\rm B}$ is accurate, as discussed earlier,
it would indicate the classical and relativistic effects are being
overwhelmed by other forces that completely reverse the direction of
net precession.

An effect that can act in such a way is a misalignment between the
spin axes and the orbital axis of the binary. This was proposed by
\cite{Shakura:1985} as an explanation for the anomalous apsidal motion
rate measured for the eclipsing binary DI\;Her, which is four times
slower than expected and had puzzled astronomers for decades.
\cite{Albrecht:2009} proved Shakura's idea to be correct by exploiting
the Rossiter-McLaughlin effect and showing that the two stars rotate
with their spin axes nearly perpendicular to the orbital axis, in such
a way as to account for the observed discrepancy \citep[see
  also][]{Claret:2010}. While a similar effect could be operating in
the 6 day binary within \vstar, it would not be sufficient to reverse
the direction of the precession, particularly since the relativistic
term dominates over the rotational terms.

An alternate possibility is dynamical interactions induced by the 2.4
day binary, especially given that the outer orbit is very eccentric
($e = 0.8533$). At closest approach the centers of mass of the two
binaries come within 1.9\;au of each other, or about 20 times the
semimajor axis of the 6 day binary.  At this distance the 2.4 day
binary may no longer appear to the other as a point source, but rather
as a larger perturbing object the size of its own semimajor axis.
Although it is not very common, normal prograde apsidal motion can be
altered drastically and even reversed by the interactions
\citep{Eggleton:2001, Borkovits:2016}, and will generally also lead to
changes in other orbital elements, an effect we have not considered
here. Examples of retrograde apsidal motion have been reported, e.g.,
by \cite{Borkovits:2015}, some as rapid as we see in \vstar.
Numerical simulations that are beyond the scope of this paper may be
able to quantify the interactions more accurately, although our
current knowledge of the quadruple system does not constrain the
problem completely. For example, we do not know how the three orbits
are oriented in space (relative inclinations), and hence their true
directions of motion, which can have a significant impact on the
perturbations. Only their line-of-sight inclinations have been
measured. They happen to be quite similar to each other ($i_{\rm A} =
83.2 \pm 1.0$\;deg, $i_{\rm B} = 85.3 \pm 1.1$\;deg, and $i_{\rm AB} =
79.6 \pm 6.4$\;deg), which might suggest near coplanarity.  Studies of
orbital alignment in hierarchical triple systems do in fact report
that relatively tight triples with outer orbits having semimajor axes
smaller than 50\;au (\vstar\ has 12.9\;au; Table~\;\ref{tab:results2})
tend to be aligned \citep{Tokovinin:2017}, although this appears to
depend also on mass, with massive systems such as \vstar\ being less
aligned than low-mass systems, on average.  If the three orbits in
\vstar\ are in fact closely aligned this would be at odds with the
known cases of retrograde apsidal motion in triple systems, which are
found to occur in strongly misaligned or even counter-rotating
configurations, driven by the Kozai-Lidov mechanism
\citep[e.g.,][]{Borkovits:2011, Borkovits:2015}.

Measuring accurate times of eclipse for the 6 day binary would be
highly beneficial to confirm or strengthen the determination of
$\dot\omega_{\rm B}$. Although the eclipses are shallow ($\sim$2.3\%),
they are well within the detection limits of many observing facilities
such as those used to search for transiting planets.

\acknowledgements

We are grateful to B.\ Beky, P.\ Berlind, Z.\ Berta, W.\ Brown,
M.\ Calkins, G.\ Esquerdo, D.\ W.\ Latham, R.\ P.\ Stefanik, and
G.\ Zhou for help in obtaining the spectroscopic observations of
\vstar, and to J.\ Mink for maintaining the CfA echelle database over
the years. We thank J.\ Irwin for helpful discussions about the use of
his light-curve code, T.\ Mazeh and M.\ Holman for conversations about
the dynamics of multiple systems, and P.\ Harmanec for help in the
initial stages of the analysis. The anonymous referee provided helpful
comments on the original manuscript. The authors also wish to thank
Dr. A.\ W.\ (Bill) Neely, who operated and maintained the NFO WebScope
for the Consortium and who handled preliminary processing of the
images and their distribution. The present analysis made use of Hydra,
the Smithsonian Institution High Performance Cluster (SI/HPC).
G.T.\ acknowledges partial support for this work from NSF grant
AST-1509375.  The research at Tennessee State University was made
possible by NSF support through grant 1039522 of the Major Research
Instrumentation Program. In addition, astronomy at Tennessee State
University is supported by the state of Tennessee through its Centers
of Excellence program. The research of M.W.\ was supported by grant
GA15-02112S from the Czech Science Foundation. This work has made use
of the SIMBAD and VizieR databases, operated at CDS, Strasbourg,
France, and of NASA's Astrophysics Data System Abstract Service.


\end{document}